\documentclass[12pt,preprint]{aastex}

\usepackage{graphicx}

\newcommand{\netsig}{$\cal N$}
\newcommand{\stellarity}{${\sc S}_2$}

\def\mb{\ifmmode {{\rm B_{435}}}\else
                ${\rm B_{435}}$\fi}
\def\mv{\ifmmode {{\rm V_{606}}}\else
                ${\rm V_{606}}$\fi}
\def\mi{\ifmmode {{\rm i_{775}}}\else
                ${\rm i_{775}}$\fi}
\def\mz{\ifmmode {{\rm z_{850}}}\else
                ${\rm z_{850}}$\fi}
\def\mJ{\ifmmode {{\rm J_{1100}}}\else
                ${\rm J_{1100}}$\fi}
\def\mH{\ifmmode {{\rm H_{1600}}}\else
                ${\rm H_{1600}}$\fi}
                
\def\mJi{\ifmmode {{\rm Ji_{1100}}}\else
                ${\rm Ji_{1100}}$\fi}
\def\mHi{\ifmmode {{\rm Hi_{1600}}}\else
                ${\rm Hi_{1600}}$\fi}
\def\mKs{\ifmmode {{\rm Ks_{2200$\ \AA$}}}\else
                ${\rm Ks_{2200\ \AA}}$\fi}

\def\mIRACa{\ifmmode {{\rm IRAC_{3600}}}\else
                ${\rm IRAC_{3600\AA}}$\fi}
\def\mIRACb{\ifmmode {{\rm IRAC_{4500}}}\else
                ${\rm IRAC_{4000\AA}}$\fi}
                
\def\ergcm2s{\ifmmode {\rm\,erg\,cm^{-2}\,s^{-1}}\else
                ${\rm\,ergs\,cm^{-2}\,s^{-1}}$\fi}

\begin{document}

\title{Spectrophotometrically Identified stars in the PEARS-N and PEARS-S fields}

\author{N. Pirzkal\altaffilmark{1,2}, A. J. Burgasser\altaffilmark{3},S. Malhotra\altaffilmark{4}, B. W. Holwerda\altaffilmark{1}, K. C. Sahu\altaffilmark{1}, J. E. Rhoads\altaffilmark{4},C. Xu\altaffilmark{5}, J. J. Bochanski\altaffilmark{6},  J. R. Walsh\altaffilmark{7}, R. A. Windhorst\altaffilmark{4}, N. P. Hathi\altaffilmark{8}, S. H., Cohen\altaffilmark{9}
}
\altaffiltext{1}{Space Telescope Science Institute, Baltimore, MD 21218, USA}
\altaffiltext{2}{Affiliated with the Space Science Depatment of the European Space Agency, ESTEC, NL-2200 AG Noordwijk,  Netherlands.}
\altaffiltext{3}{Massachusetts Institute of Technology, Kavli Institute for Astrophysics and Space Research, Building37, Room664B, 77 Massachusetts Avenue, Cambridge, MA02139, USA}
\altaffiltext{4}{School of Earth and Space Exploration, Arizona State University, Tempe, AZ 85287, USA}
\altaffiltext{5}{Shanghai Institute of Technical Physics, 200083, Shanghai, China.}
\altaffiltext{6}{Astronomy Department, University of Washington, Box 351580, Seattle, WA 98195, USA}
\altaffiltext{7}{ESO/ST-ECF, Karl-Schwarschild-Strasse 2, D-85748, Garching bei M\"{u}nchen, Germany}
\altaffiltext{8}{Department of Physics and Astronomy, University of California, Riverside, CA 92521}
\altaffiltext{9}{Department of Physics, Arizona State University, Tempe, AZ 85287}

\begin{abstract}
Deep ACS slitless grism observations and identification of stellar sources
are presented within the Great Observatories Origins Deep Survey
(GOODS) North and South fields which were obtained in the Probing
Evolution And Reionization Spectroscopically (PEARS) program.  It is
demonstrated that even low resolution spectra can be a very powerful means
to identify stars in the field, especially low mass stars with stellar types
M0 and later. The PEARS fields lay within the larger GOODS fields, and we
used new, deeper images to further refine the selection of stars in the PEARS
field, down to a magnitude of $\mz = 25$ using a newly developed stellarity
parameter. The total number of stars with reliable spectroscopic and
morphological identification was
95 and 108 in the north and south fields respectively. The sample of
spectroscopically identified stars allows constraints to be set on the thickness
of the Galactic thin disk as well as contributions from a thick disk and a halo
component. We derive a thin disk scale height, as traced by the population of
M4 to M9 dwarfs along two independent lines of sight, of ${\rm h_{thin} = 370^{+60}_{-65}\ pc}$. When
including the more massive M0 to M4 dwarf population, we derive ${\rm h_{thin} = 300 \pm 70pc}$. In both cases, we observe that 
we must include a combination of thick and halo components in our models in order to account for 
the observed numbers of faint dwarfs. The required thick disk scale height is typically ${\rm h_{thick}=1000 pc}$ and the
acceptable relative stellar densities of the thin disk to thick disk and the thin disk to halo components are in the range of ${\rm 0.00025<f_{halo}<0.0005}$ and ${\rm 0.05<f_{thick}<0.08}$ and are somewhat dependent on whether the more massive M0 to M4 dwarfs are included in our sample.
\end{abstract}

\section{Introduction}
The GOODS northern (GOODS-N) and southern (GOODS-S) fields \citep{giavalisco} have been observed over a wide range of wavelength and down to very faint magnitudes during the past few years. As part of the PEARS project (PI:Malhotra), we have observed 9 fields within the GOODS regions, each 11.65 ${\rm arcmin^2\ }$. These new observations provide  spectroscopic data for more than 10,000 sources, and cover a wavelength range of 6000\AA\ to 9500\AA   (${\rm 40\AA\ per\ pixel\ or\ R\approx100}$) . Ultimately, the effective resolution of these slitless spectra is determined by the physical size of each source, as projected on the sky. However, the  ACS point spread function is small (${\rm \approx 1.5\ pixels}$), and well-sampled and the maximum resolution is thus achieved when observing point sources. Such spectra of point sources, as we will show in this paper, are an excellent means to spectroscopically identify stars, especially old, low mass stars (M and later dwarfs) that have prominent and broad absorption features.

By studying the older stellar population content of our galaxy down to the faintest possible magnitude, we can directly observe and measure the shape of the thin Galactic disk as traced by the old stellar population. In \citet[][Paper I]{pirzkal2005}, and using deep slitless observations of the Hubble Ultra Deep Field (HUDF), the observed number and brightness distribution of M4 and later stellar dwarfs made it possible to directly infer that the Galactic disk had a scale height of $400 \pm 100 {\rm pc}$ for M and L stars. The relatively small fraction of the sky observed by the HUDF, however, resulted in the detection of only a handful of M dwarfs, even though the previous result from  \cite{ryan2005} was confirmed.

The study presented here allows us to include a much larger number of stars as well as performing these measurements along different lines of sights, both pointing above and below the Galactic plane.  \citet{stanway2008} used the publicly available GOODS 1.0 public data release to morphologically and photometrically select M dwarfs in those fields. The authors claim to see no decrease in the number counts of these objects down to \mz=26 and moreover observed a perplexing 34\% overabundance of these objects between the GOODS-N and GOODS-S fields.

In this paper, we present a new analysis of the stellar content of a large fraction of the GOODS-N and GOODS-S fields, using GOOD 2.0 public data (\url{http://archive.stsci.edu/prepds/goods/}). Section \ref{description} briefly describes the observations and data we used for this work. Our initial morphological selection and spectroscopic fitting are described in Section \ref{preselect}, while the refined selection of stellar candidates is discussed in Section \ref{star_selection}. We discuss spectroscopically identified M and later dwarfs, their color, magnitude and distance distribution as well as the implication these have on the possible thickness of the Galactic disk in Section \ref{MD}.

\section{Observations}\label{description}
PEARS observations were obtained as part of a large HST proposal (200 orbits, Proposal 10530, PI: Malhotra) that was closely modeled after the previously very successful GRAPES \citep{pirzkal2004} observations of the much smaller subset of the GOODS-S field: the Hubble Ultra Deep Field \citep{beckwith}. Each of the 9 PEARS fields (11.65 ${\rm arcmin^2\ }$ each) was observed for approximately $\approx 40000$ seconds (20 orbits), split evenly between observations taken at different position angles on the sky (typically 3). One of the five PEARS-S fields contains additional observations of the previous GRAPES/HUDF field. Of the eight remaining observed fields, four are within GOODS-N and four are within GOODS-S. The combined areas of the PEARS-N and PEARS-S fields are 50.17 and 70.61 ${\rm arcmin^2\ }$, respectively.
The PEARS data were reduced using the latest version of the ACS slitless extraction program, aXe \citep[\url{http://www.stecf.org/software/slitless\_software/axe/}]{kummel2008}, following the recipe described in \citet{pirzkal2004}. A few aspects of the data reduction process differ somewhat however: First. the background subtraction was improved, and the spectra were extracted using optimal extraction. More significantly, the amount of contamination, caused by overlapping spectra of nearby sources was quantitatively estimated for each spectrum. These changes allowed us to reach slightly higher signal-to-noise levels at any given broad band magnitudes than for the GRAPES project. The PEARS data reduction is described in more detail in a companion paper by \cite{malhotra2008}. The extraction was based on an object catalog that, crucially, was derived using newly available and deeper broad band images of the GOODS fields. These observations, known as GOODS 2.0 combine the older GOODS observations with additional ACS \mi\ and \mz\ band images  subsequently obtained as part of a large supernova search program (PI: Riess). A total of 4081 and 5486 spectra were extracted from the PEARS-N and PEARS-S fields, respectively. The broad band magnitudes quoted in this paper are AB magnitudes and were based on our own Sextractor \citep{Sextractor} generated catalogs of the ACS \mb, \mv, \mi\ and \mz\ band images. As a general rule, spectra obtained at different position angles were not combined together. Instead, specific fitting and measurements were made separately and then averaged, allowing us to estimate the error in the fit or measurements by computing the standard deviation of the mean.

\section{Object selection}\label{preselect}
\subsection{Morphological pre-selection}
\begin{figure}[h]
\includegraphics[width=3.0in]{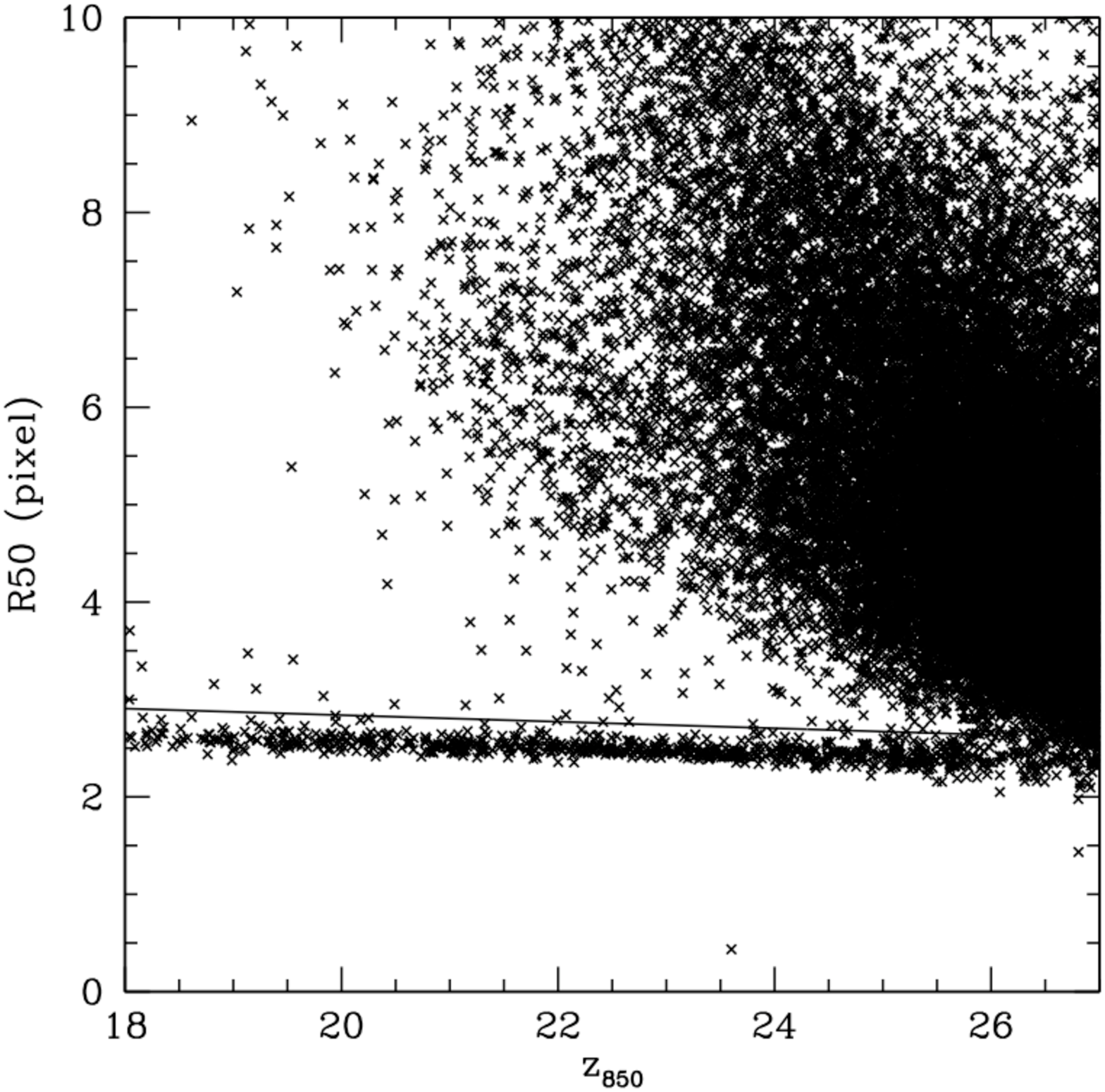}
\includegraphics[width=3.0in]{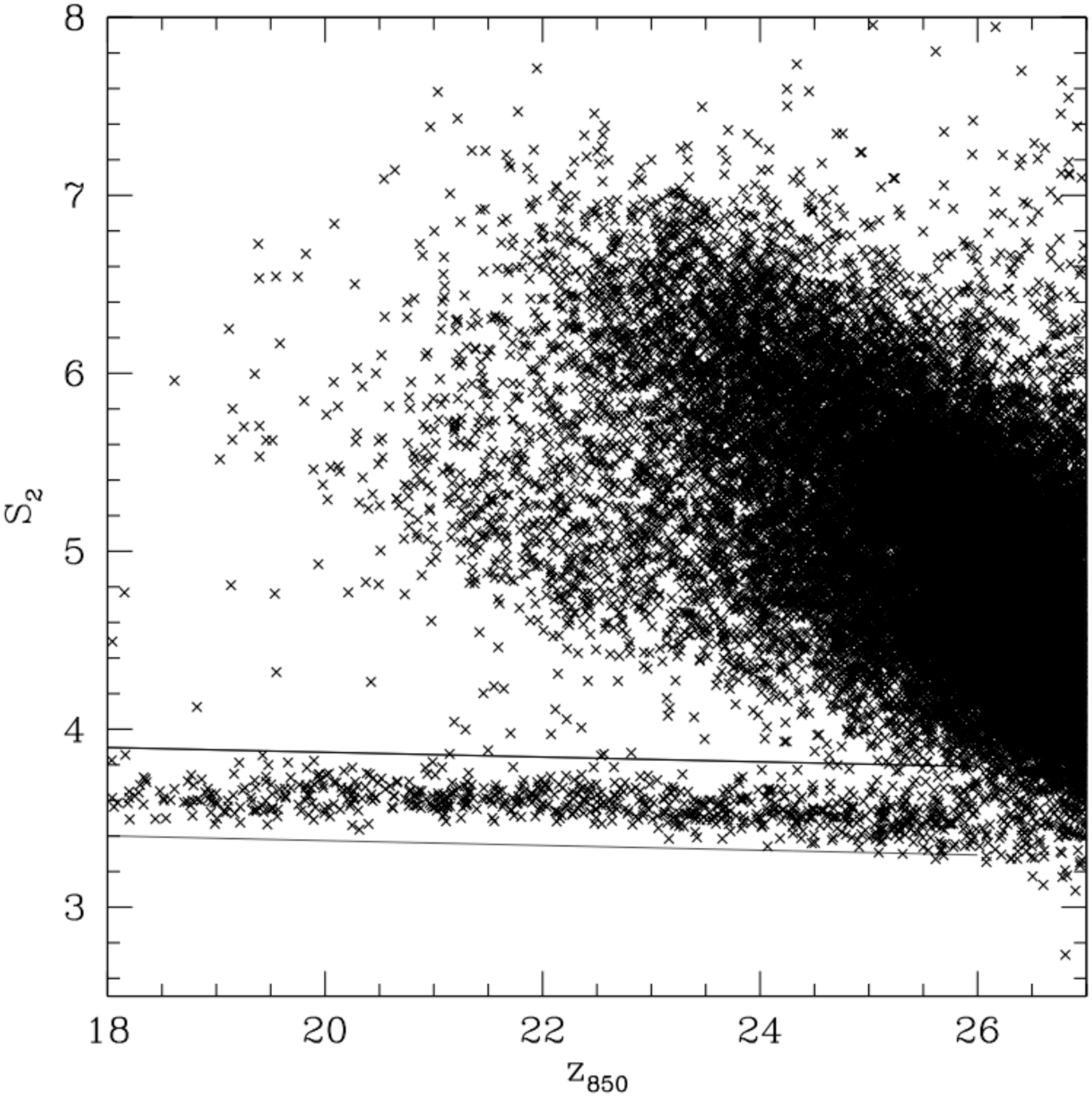}
\includegraphics[width=3.0in]{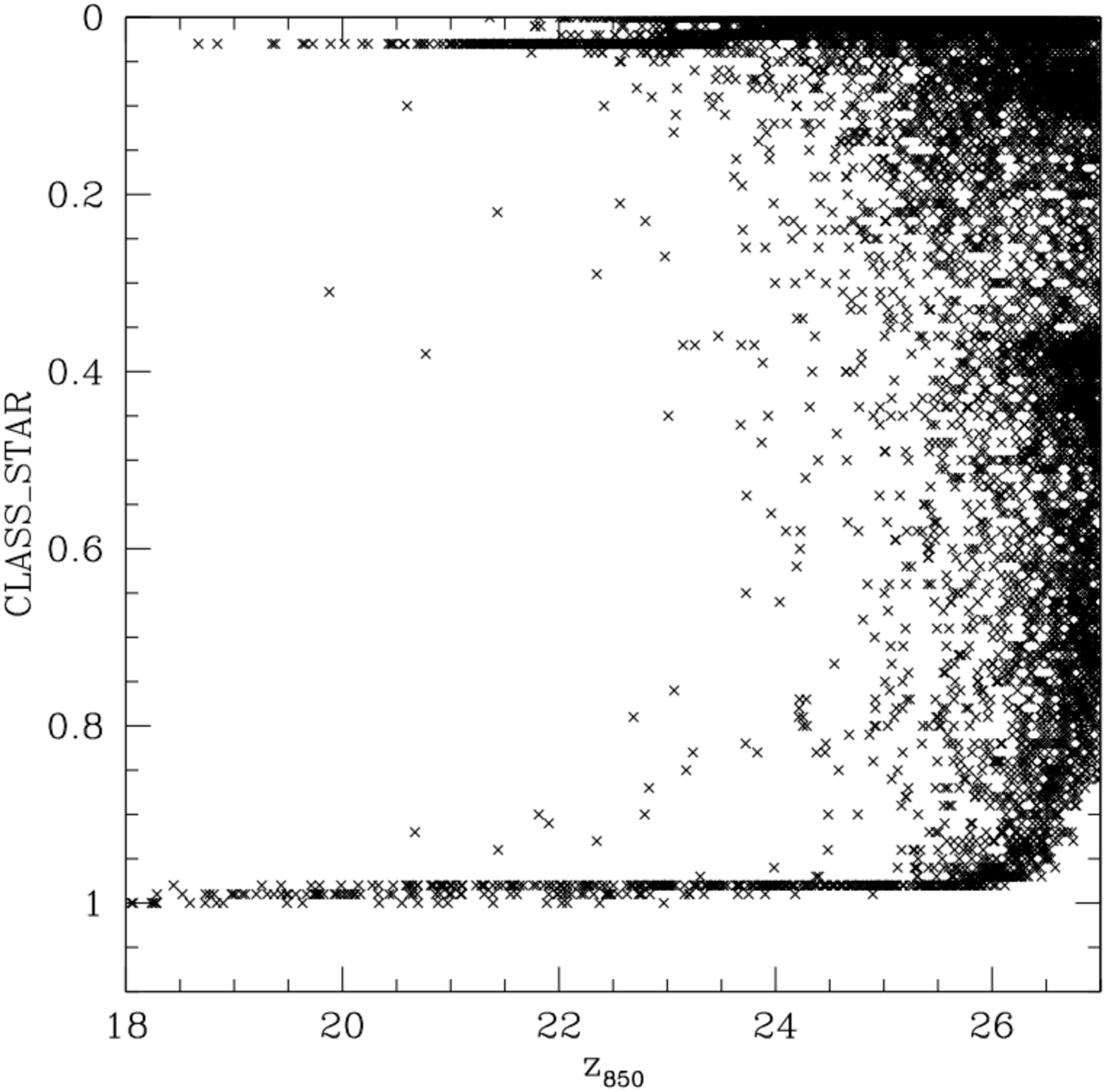}
\caption{\label{R50} The half-light radii, stellarity \stellarity, and Sextractor CLASS\_STAR parameter in the \mz\ band for all objects in the PEARS fields.}
\end{figure}

Traditionally, stars can be distinguished from extra Galactic objects because they are unresolved. However, deeper surveys such as the HUDF or GOODS reach down to magnitudes faint enough that the faintest galaxies in the field are not resolved and are hard to distinguish from stars. For example, Paper I, \citet{kilic2005}, and \cite{stanway2008} have used different morphologically based selection criteria to identify unresolved objects in deep fields. All of these methods have been shown to work relatively well, especially for relatively bright objects ($\mz < 24$),  but they all tend to progressively fail at fainter magnitudes, where one can no longer differentiate between stars and extraGalactic objects using only their observed sizes. As seen in Figure \ref{R50}, the population of stars and resolved sources can be reasonably well separated down to magnitudes as faint as $\mz \approx 26$ using a variety of parameters, including half-light radius (${\rm R_{50}}$), Sextractor CLASS\_STAR, and a new parameter which we refer to as the stellarity parameter (\stellarity). The latter is defined as the ratio of the flux within a circular aperture of radius 1 pixel, centered on the second moment position of the source, to the total isophotal flux as measured by Sextractor. It is computed as  (MAG\_APER - MAG\_ISO) as measured using  \mz\ band GOODS 2.0 images. 

As we already pointed out above, however, all of these selection criteria have the tendency to fail in different ways at the faint end. It is therefore difficult to come up with a clear morphological selection of stellar objects at the faint end. Instead of somewhat arbitrarily setting acceptable upper values for any of these parameters,  we instead levied the available deep broad band imaging with the deep low resolution spectra of the sources in these fields. While we initially simply started by systematically fitting the spectrum of nearly every source in the field by comparing them to a set of standard templates (Section \ref{fitting}), we refined the selection process by identifying valid values of \stellarity\  for robustly spectroscopically identified stars (Section \ref{star_selection}). 

\subsection{Spectroscopic fitting}\label{fitting}
As a first step, we fitted the spectra of all sources,  excluding only sources that are grossly extended. Thus,  all objects with  ${\rm CLASS\_STAR > 0.1}$ or ${\rm R_{50} < 50.0\ pixel}$, and $\mi < 28.$ (38\% of all the available sources) were spectroscopically fitted. 
Spectral fitting was performed using a set of stellar templates from \citet{pickles1998} for A to K type stars, from \citet{bochanski2007} for M dwarfs, and \citet{Kirkpatrick1999} for L and T dwarfs. We first simulated point sources using these templates, generating realistic two dimensional images of the dispersed spectra, and re-extracted these using the same method that we used to extract the PEARS data. This allowed us to produce realistic stellar templates with appropriate resolution and all the known characteristics of ACS slitless grism observations. We then fitted PEARS spectra to every one of these smoothed templates, in e-/s space, minimizing the $\chi^2$ computed between observed spectra and templates while only allowing for a scaling factor to be varied, as well as a small shift in the wavelength dispersion of at most a few pixels. The latter was implemented in order to account for potential errors in the estimate of the spectral zero point in each PEARS spectra which could result in a slight shifting of the observed spectra in the wavelength direction. 

\begin{figure}[h]
\includegraphics[width=6.0in]{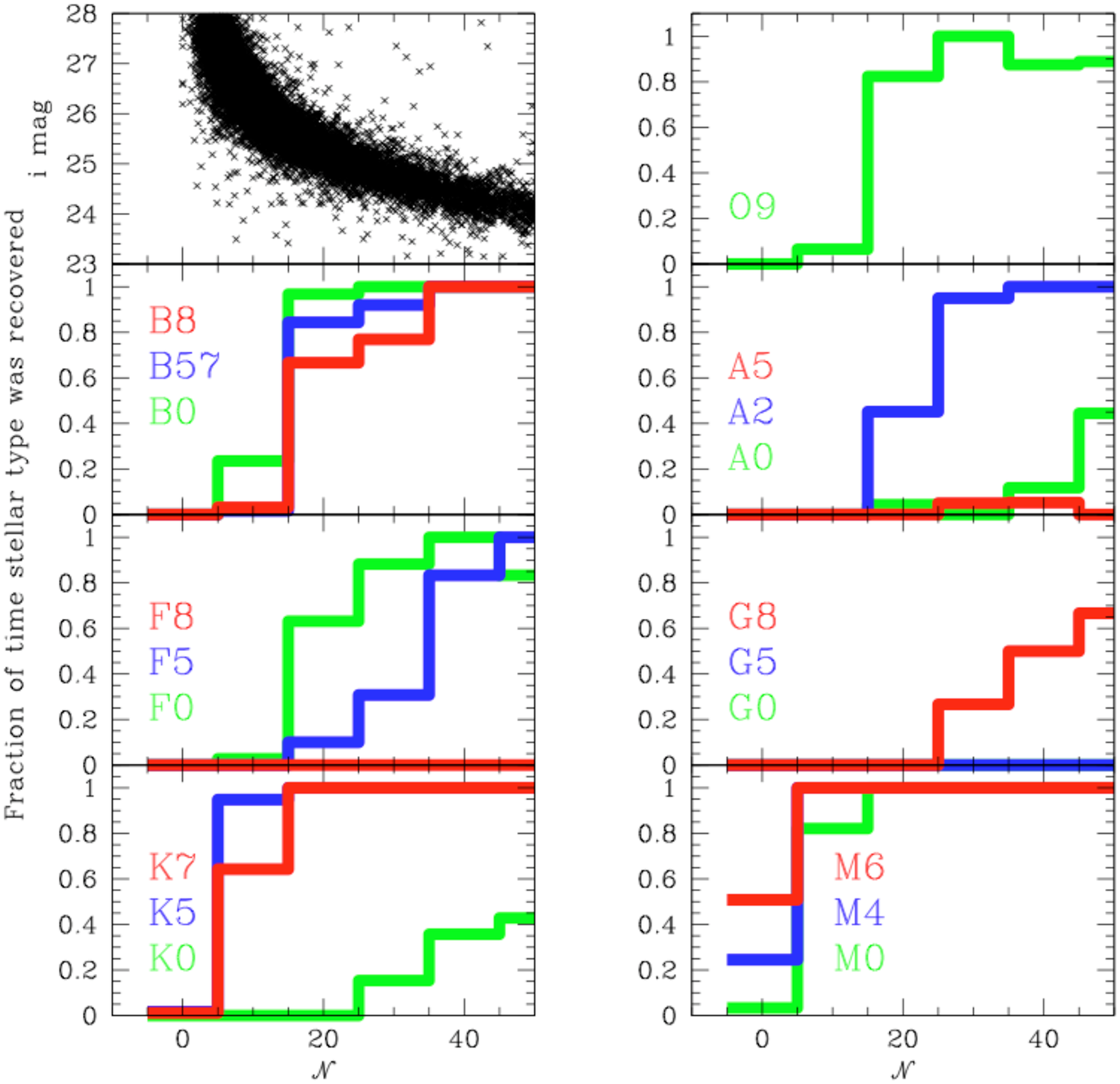}
\caption{\label{fracAll01} This figure shows the fraction of time that un-ambiguous spectroscopic stellar determination was possible, via our fitting process, to simulated spectra of various spectral types.  From top to bottom and left to right, we show our results for O, B, A, F, G, K, and M stars. These plots show the fraction of occurences that an object was assigned the proper stellar type with a standard deviation in the assigned stellar type smaller or equal to 0.1 sub-type (e.g. M4-M5). These are plotted as a function of \netsig, the cumulative signal-to-noise in the input simulated spectra (Section \ref{fitting}). The top left quadrant shows the relation between measured \netsig and \mi magnitude for the entire PEARS sample. As shown here, a \netsig of 20 corresponds to $\mi \approx 25$ and  a \netsig of 10 to $\mi \approx 26$}
\end{figure}

In order to quantify the quality of individual PEARS spectra, we re-introduce the Netsig parameter (\netsig) described in \citet{pirzkal2004}. \netsig\ is defined as the 
maximum cumulative signal-to-noise level. It is computed by sorting extracted spectral bins in a spectrum by decreasing signal-to-noise. Iteratively, an increasing number of sorted bins are then added up together, while their associated noise is added in quadrature,  to compute the signal-to-noise in the accumulated signal. This iterative process is stopped when the signal-to-noise in the accumulated signal (\netsig) no longer increases when an additional bin is considered. A high value of \netsig\ can be reached either by an object with a faint, but significant level of continuum and no emission line(s), or by an object possessing prominent emission lines but with no detectable continuum, or a combination of detected continuum and emission line(s). Note that a value of \netsig$ \approx 10$  corresponds to an object with a continuum \mi\ band magnitude of about 26. The \netsig\ values were computed for all of the PEARS spectra and these are shown as a function of \mi\  magnitude in the topmost left panel of Figure \ref{fracAll01}.

In order to quantify our ability to spectroscopically identify stars using PEARS spectra using our set of spectral templates, we generated a series of $\approx 50,000$ simulated stellar spectra, using the aXe software to both simulate and extract the simulated spectra. The stellar type was randomly chosen, and the broad band magnitude of the objects were allowed to vary from $22 < \mz < 28$, resulting in a large range of \netsig\ values. We fitted the simulated spectra to our stellar templates and simply kept track of the fraction of time that our fitting routine determined the proper stellar type (within 1 subtype) as a function of input stellar type and \netsig\ values. The result of these simulations is summarized in Figure \ref{fracAll01}. The latter clearly shows  that, if the templates are proper and accurate representations of the various spectral types, we are indeed able to reliably identify most stars spectroscopically. While A, F, and G stars, objects that have a relatively flat and featureless color gradient over the spectral range of the grism, are only properly spectroscopically identified when bright ($\netsig > 40$), K and later type stars are well identified down to faint magnitudes and small values of \netsig. In particular, using PEARS spectra, we can spectroscopically identify M0 and later stars down to low value of  \netsig$=10$ ($\mz \approx 26$ for a pure continuum object as shown by the top left panel of Figure \ref{fracAll01}).

\begin{figure}[h]
\includegraphics[width=3in]{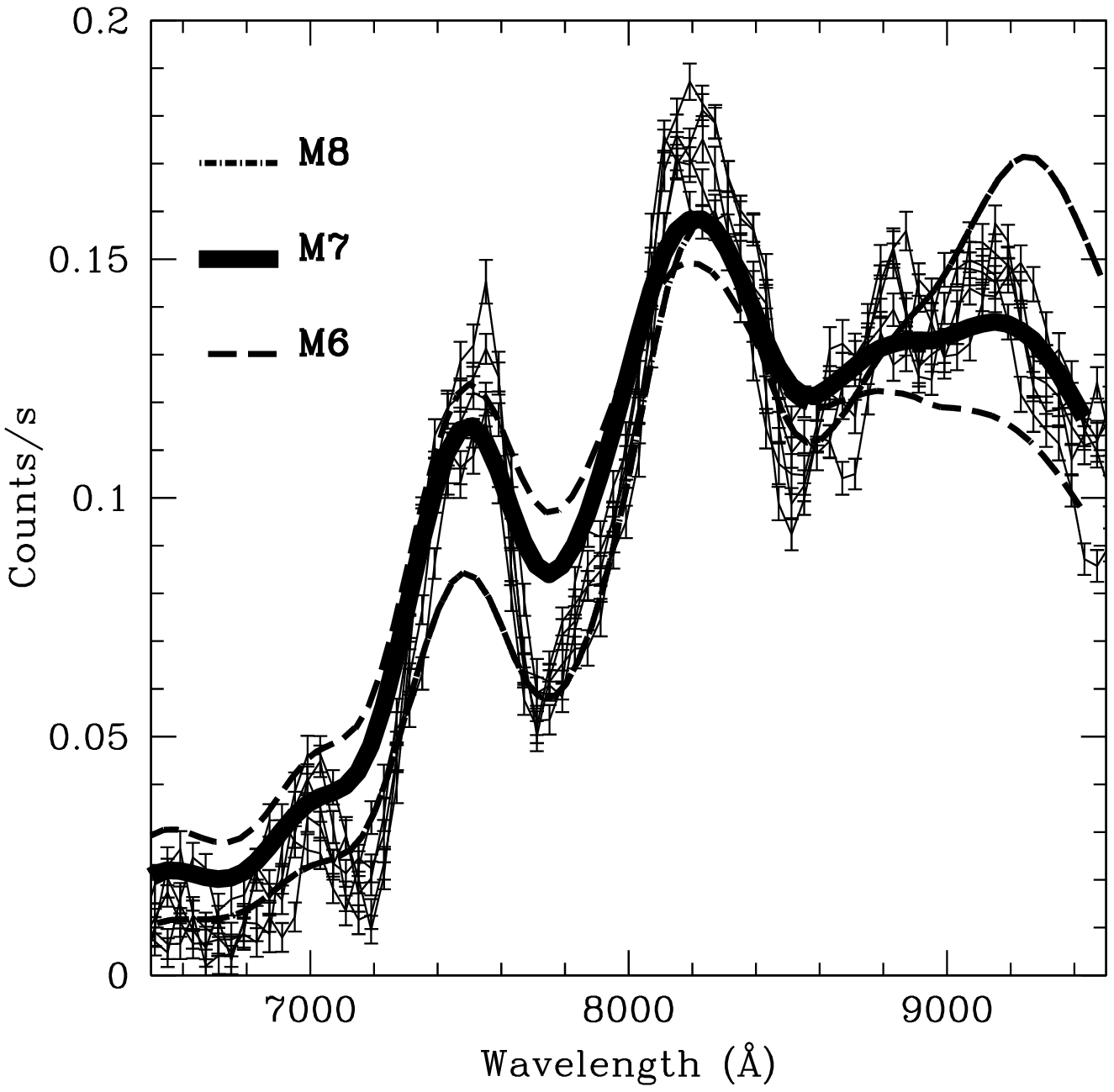}
\includegraphics[width=3in]{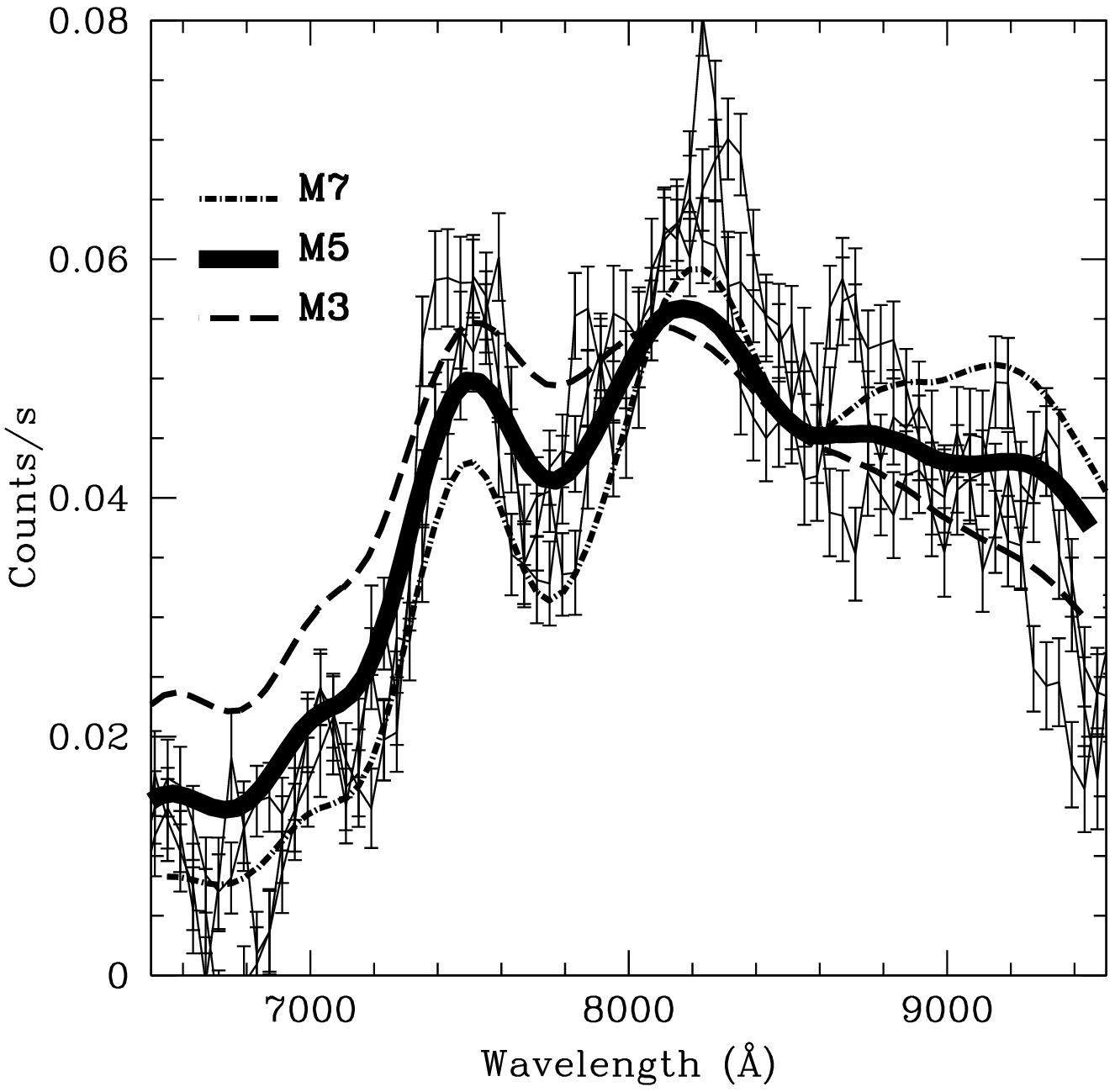}

\caption{\label{M7_example} The leftmost plot shows 5 distinct observations of a \mz=22.9 stellar candidate (overplotted using thin lines and error bars). Overplotted is the best spectra fit to this object M7 stellar template from \citet{bochanski2007} as well as M6 and M8 templates (dash lines). The rightmost plot shows a fainter \mz=24.2 object best fitted by an M5 template. The M3 and M7 best fits are also shown.}
\end{figure}

However, and while this demonstrates that even a low resolution spectrum can be a very powerful tool to determine the type of such objects, the reality is that the input templates from \cite{pickles1998} and  \citet{bochanski2007} are not always a perfect fit to the observations and that small variation in the inherent nature of these objects can cause significant and systematic variations, as shown in Figure \ref{M7_example}. While there is little doubt that the type of the object is properly assigned, there are systematic differences between the template and the high signal-to-noise observed spectrum which cause the $\chi^2$ of the fit to be higher than expected. The implication is that the $\chi^2$ fit, while appropriate to find the template that best fit an observation, cannot be used as a measure of the quality of the fit alone. This is particularly important in brighter objects where the main source of uncertainty is not random and caused by photon statistics but is instead systematic error in the templates that are used. Photon noise becomes dominant for the dimmer objects and we observe that for sources fainter than $\mz \approx 23$ the $\chi^2$ statistics can again be used to estimate the quality of the fit. In any case, and for  brighter PEARS sources, we therefore resorted to individually examining each of the PEARS spectra and their best fitted templates to ensure that these were good fits. Since each of PEARS sub-fields were observed separately using different position angles, we are able to check the consistency and robustness of an assigned stellar type by computing the variance between the types of the best fit templates at different position angles. We find that this variance is typically better than one sub-type (e.g. M4-M5) for all M dwarfs brighter than \mz=25. Objects observed only once were not considered in this study. For all sources (with any spectral type) brighter than $\mz \approx 23$, as mentioned above, we visually ensured that the spectral fits were good. While the $chi^2$ for bright M dwarfs were sometimes higher than expected, as we explained above, we found the spectral identification of these sources to be very good and did not have to correct or remove any M dwarfs from our sample. For all fainter sources we considered a fit to be good if $\chi^2 < 2$, with the exception of late K and M dwarfs candidates which we individually checked to be well fit. 

%
\section{Selecting stars}\label{star_selection}
\begin{figure}[h]
\includegraphics[width=5.0in]{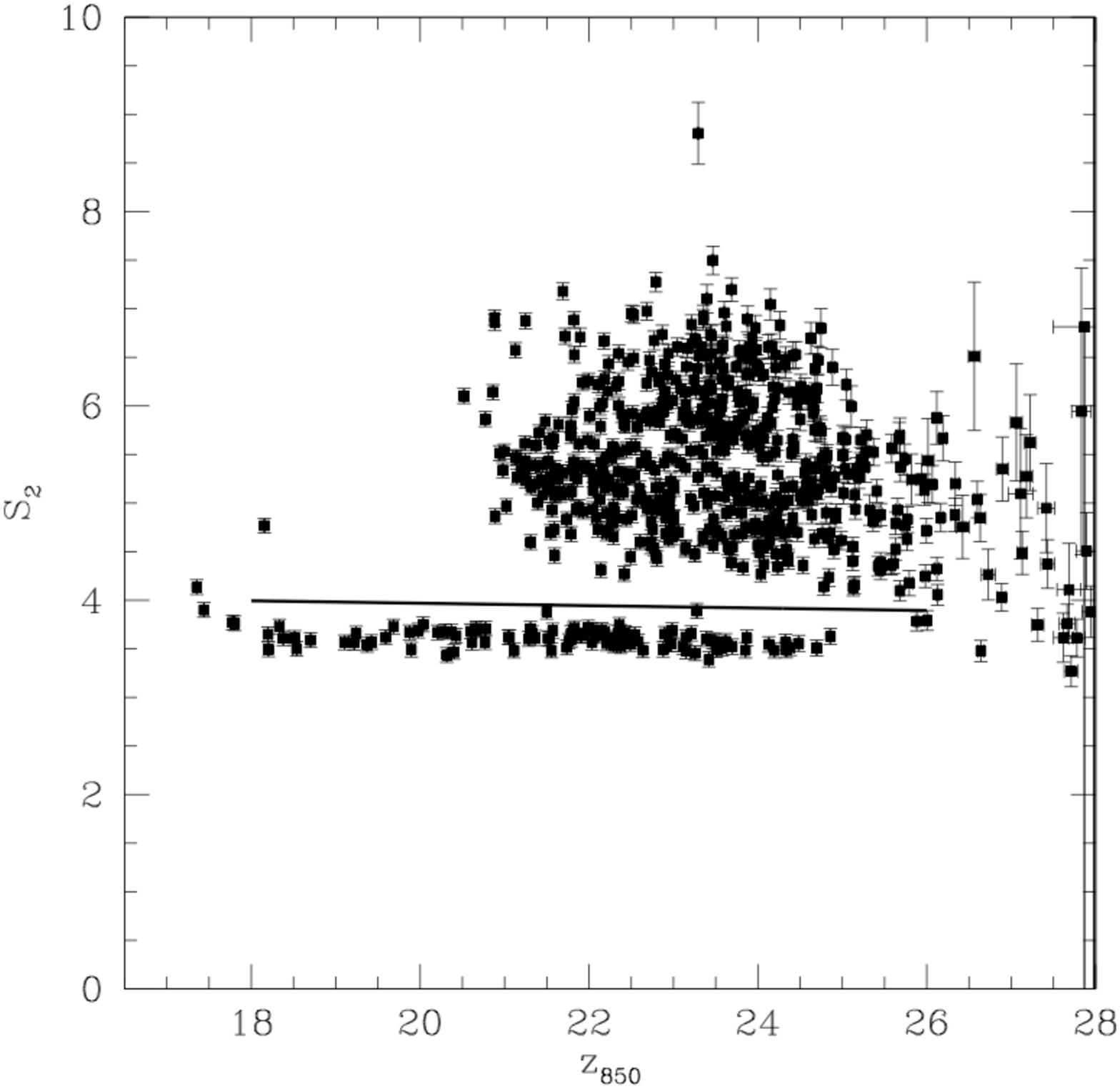}
\caption{\label{M1_later_select} Stellarity, \stellarity, of objects that are fitted best by an M1 or later stellar template spectrum. The distribution of stars is well separated from other objects in the field. The nearly horizontal line shows our selection criteria for unresolved objects. Objects below this line are stars and objects above this line are red galaxies whose closest stellar spectral match are M1 and later stars, but are not necessarily well fitted by them. It is defined as a $3 \sigma$ (standard deviation of the mean) envelope to the observed locus of M1 and later dwarfs.}
\end{figure}

In Section \ref{fitting}, we described a first pass at spectroscopically fitting every source in the field. However, this brute force method failed to levy the deep, high resolution (0.03'' per pixel) GOODS 2.0 images together with the information contained in our PEARS spectra. As mentioned in Section \ref{preselect}, identifying stars morphologically can work well but particular attention must be paid when selecting faint objects. Instead of simply basing our selection of stellar sources on Figure \ref{R50}, we can fine tune our selection using values of  \stellarity. We already saw how the PEARS spectra allowed us to easily identify low mass M and later dwarfs in the field, and we therefore choose these objects to carefully define acceptable values of \stellarity\ to morphologically identify stars with all stellar types in the field. We started by rejecting any object that was not observed at more than one position angle in the PEARS data. We furthermore rejected any object for which the best fitting template varies, from one observation to the next, by more than one stellar sub-type. Finally, we rejected objects fitted by templates with stellar types earlier than M1. The result of this selection, showing only objects reasonably best fitted by stellar spectra of stellar types M1 and later, is shown in Figure \ref{M1_later_select}. Not all of these objects, galaxies for example, are necessarily well fitted by stellar templates to start with,  but these are objects which, when fitted only to stellar templates, resemble most our M1 or latter stellar templates. In this figure, showing \stellarity\ versus \mz, the locus of stars is clearly visible and separated from that of field galaxies.  Individual objects within this locus of stars were individually checked (both morphologically and spectroscopically) and this set of stars, which were hence spectroscopically identified, was subsequently used to redefine our morphological selection. 

\begin{figure}[h]
\includegraphics[width=5.0in]{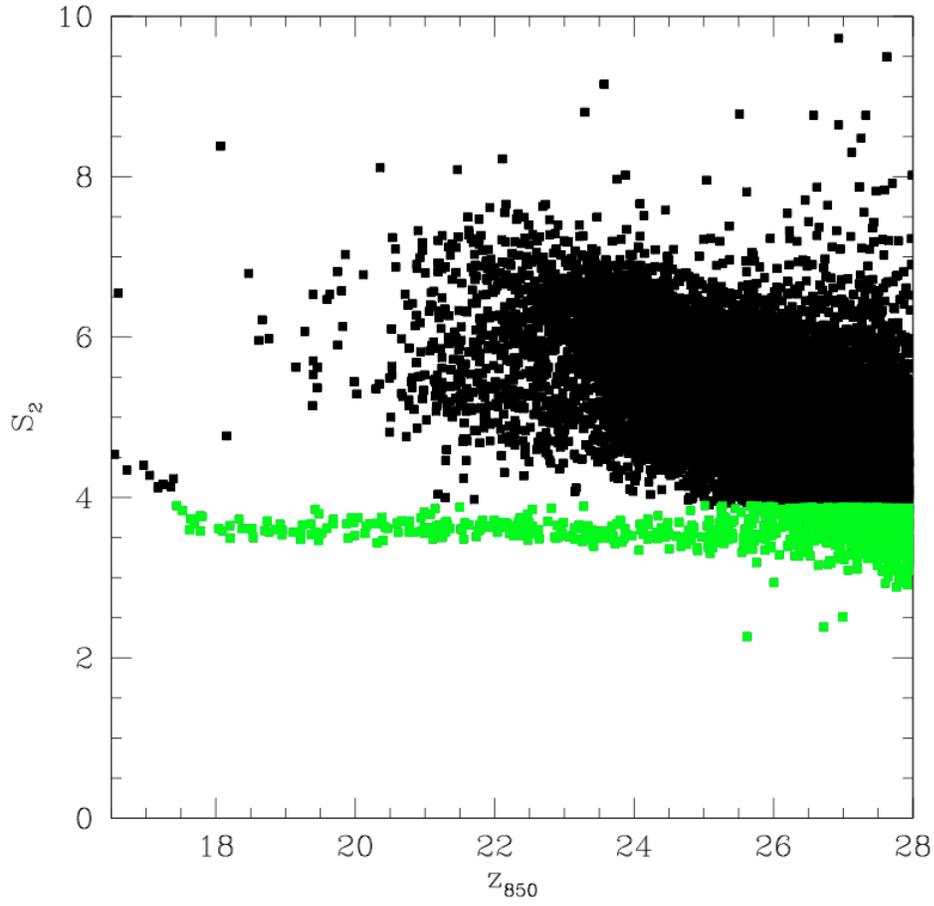}
\caption{\label{star_select} Stellarity, \stellarity, of objects in PEARS and the subset of unresolved sources shown in green. Objects with larger stellarity at magnitudes  brighter than $\mz \approx 17.5$\  are saturated objects.}
\end{figure}

As shown by the horizontal line in Figure  \ref{M1_later_select}, the acceptable values of the \stellarity\ as a function of magnitude for the known spectroscopically identified stars in the fields are  \stellarity$<4.22 - 0.0125 \times${\mz}, where \mz\ is the MAG\_ISO Sextractor measured quantity. The latter serves as our morphological selection criterion for the rest of this study. Note that we relied on the \stellarity\ instead of the somewhat more familiar half-light radius parameter (${\rm R_{50}}$) simply because, while both methods identified the same stars in the fields (down to $\mz \approx 26.5$), we found that selecting objects based on values of ${\rm R_{50}}$ resulted in the selection of a handful of  extra-Galactic interlopers, often simply small galaxies with  complex structures. 
As the final step of our stellar selection process,  we applied our \stellarity\ based criterion to the entire sample of objects in the PEARS fields, independently of any spectroscopic fitting (hence allowing us to include stars that might not be so easily identified spectroscopically, namely those with a flatter spectra energy distribution, i.e. A to K type stars). 
Figure \ref{star_select} shows  the sub-sample of the PEARS sources that were thus selected. This Figure also shows that for $\mz > 26$, the locus of unresolved extra-Galactic objects begins to contaminate the number of stars in the field. We therefore conservatively restricted our morphological selection of stellar candidates to only objects brighter than $\mz = 26$. The resulting number of stellar candidates are 129 and 154 in PEARS-N and PEARS-S respectively, for a total of 283 in both fields. Down to $\mz = 25$, a conservative limit to our ability to spectroscopically identify stars, the number of sources is 95 and 108, respectively, for a total of 203 sources, Note that we are not assuming that galaxies cannot have spectra that are well matched by stellar spectra, but that these would either be close enough to be resolved or far enough to be redshifted. In either case, such objects would be either rejected by our morphological selection or a poor spectral match to our non-redshifted stellar templates.

\begin{figure}[h]
\includegraphics[width=5.0in]{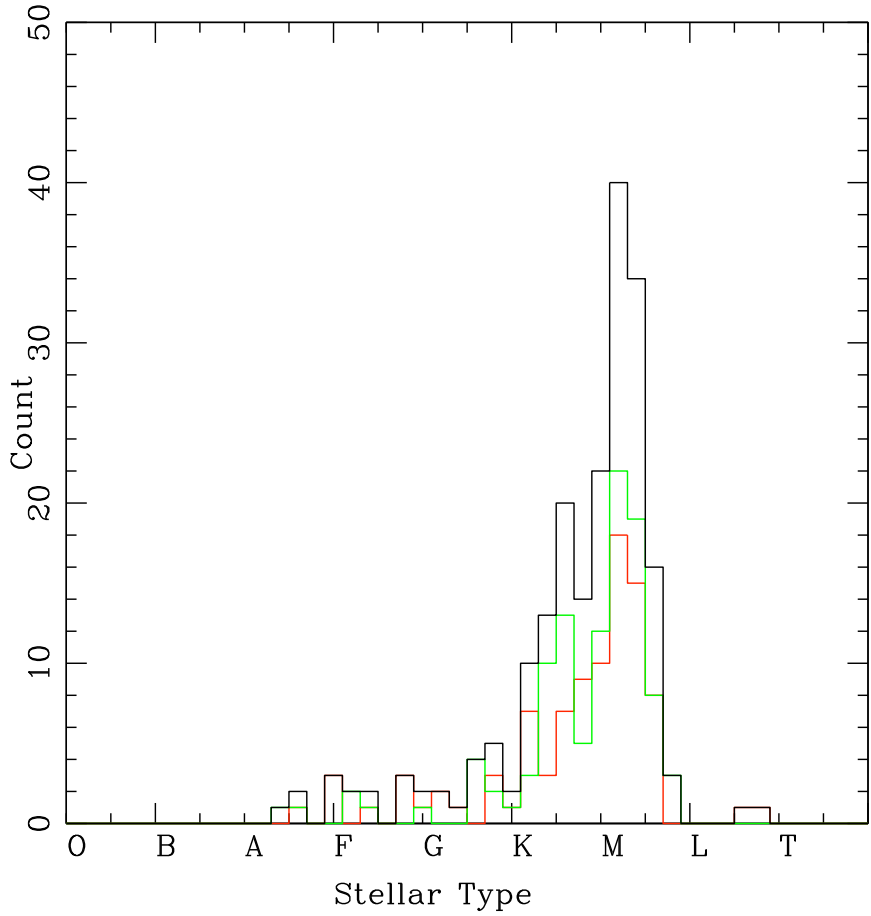}
\caption{\label{star_select_hist} Number of stars identified as a function of stellar type for sources brighter than \mz=25. Each bin is 0.2 of a stellar type wide. PEARS-N and PEARS-S are shown in green, respectively. The combined PEARS fields are shown in black.}
\end{figure}

\begin{figure}[h]
\includegraphics[width=5.0in]{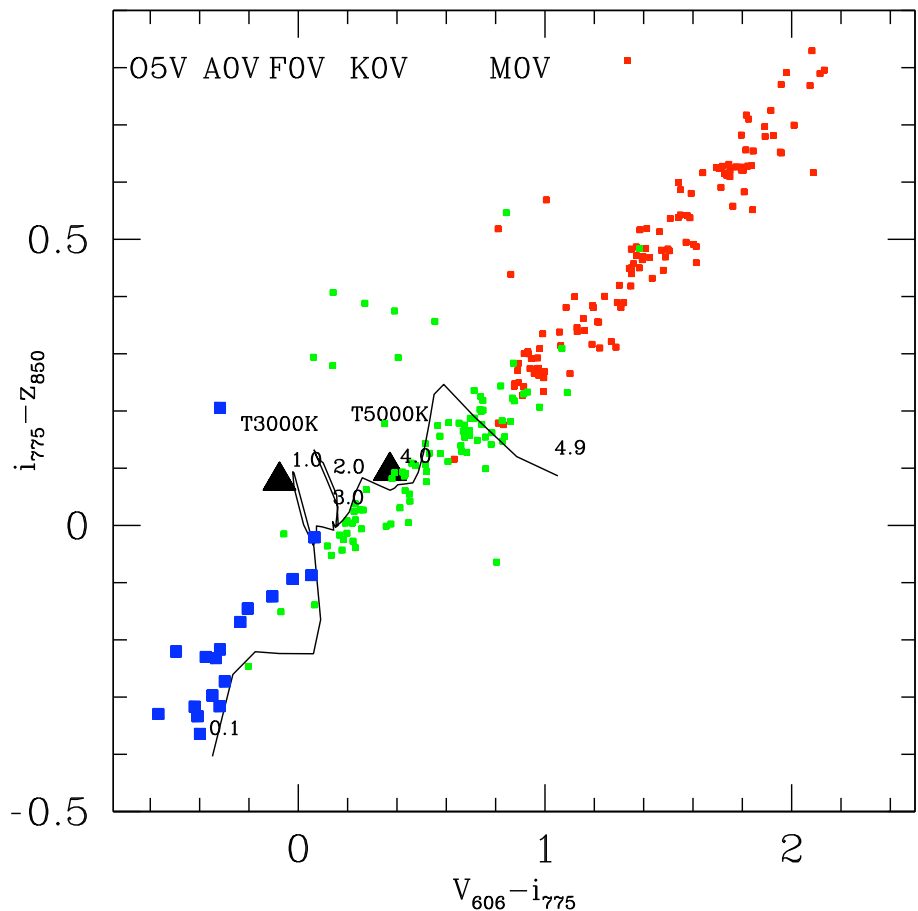}
\caption{\label{izvi} This plot shows the PEARS M0 and later stars (red) and others (green) in a color-color plot, together with cool  white dwarfs \citep[3000K and 5000K shown with solid triangles,][]{harris2001}. Hot white dwarfs are shown using solid blue squares. The thin black line shows the position of QSO's from redshift z=4.9 (top right) to 0.1 (bottom left).
}
\end{figure}

In Figure  \ref{star_select_hist} , we present a histogram of the spectral type of the PEARS stars down to $\mz=25$, keeping in mind however that our ability to accurately determine the type of a star is itself stellar type dependent and is most reliable only for late type stars (and very early ones). As shown in this plot, we only identify a handful of A or earlier stars in the field and the main fraction of the stars are identified as K or later types, where our spectroscopic identification is robust. We show the entire sample of PEARS stars in a color-color plot in Figure \ref{izvi}.

\begin{figure}[h]
\includegraphics[width=5.0in]{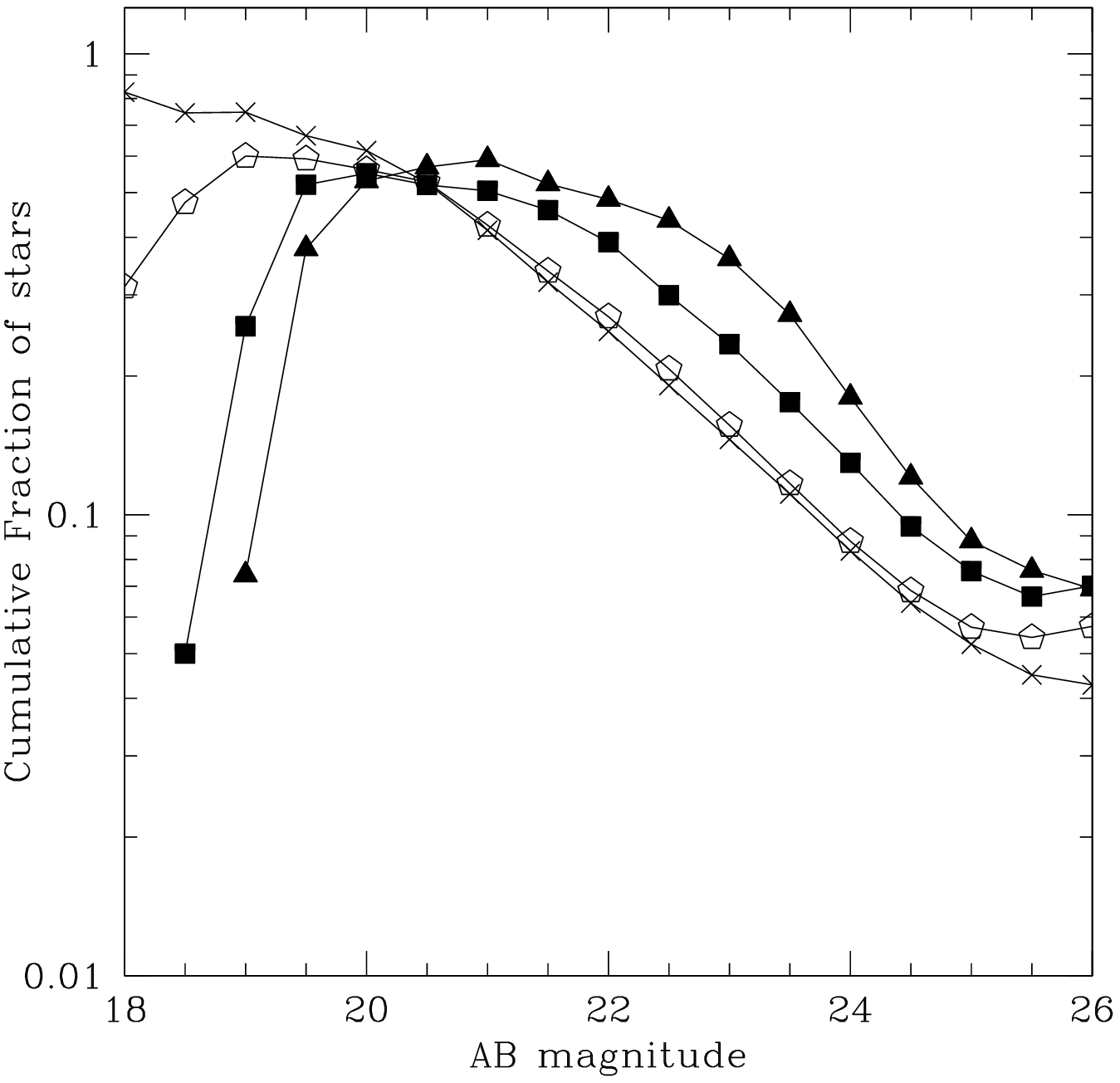}
\caption{\label{stars_gals} Cumulative fraction of stars in the PEARS fields as a function of magnitude in the \mb\ (filled triangles), \mv\ (filled squares), \mi\ (open pentagons) and \mz\ (crosses) bands, respectively. This figure shows that object number counts in the z band are clearly dominated by stars at $mz < 22$ and remain a significant source of contamination in all bands down the limiting magnitudes of the GOODS and PEARS surveys.}
\end{figure}

At this stage, it is worth addressing the issue of stellar contamination in deep extra-Galactic studies such as GOODS. In Figure \ref{stars_gals} we show the ratio of the cumulative number of stars to the cumulative number of extra-Galactic sources as a function of \mz\ magnitude. These cumulative distributions in the four ACS bandpasses were derived directly by from Figure \ref{star_select}. As shown in Figure \ref{stars_gals}, if we consider the cumulative number of objects down to a specific limiting magnitude, an object catalog of a field such as the PEARS/GOODS fields will contain a significant fraction of stars. In the reddest  band ($z_{850}$) in particular, the majority of objects brighter than $\mz \approx 21.5$ are stars while at the fainter magnitudes of $\mz > 25$, 4\% of objects remains stars. As shown in Figure \ref{stars_gals}, the effect is subtly different in each of the ACS bandpasses with the bluer bands being less dominated at magnitudes $\mz < 20$.

\section{M dwarfs}\label{MD}
\begin{figure}[h]
\includegraphics[width=5.0in]{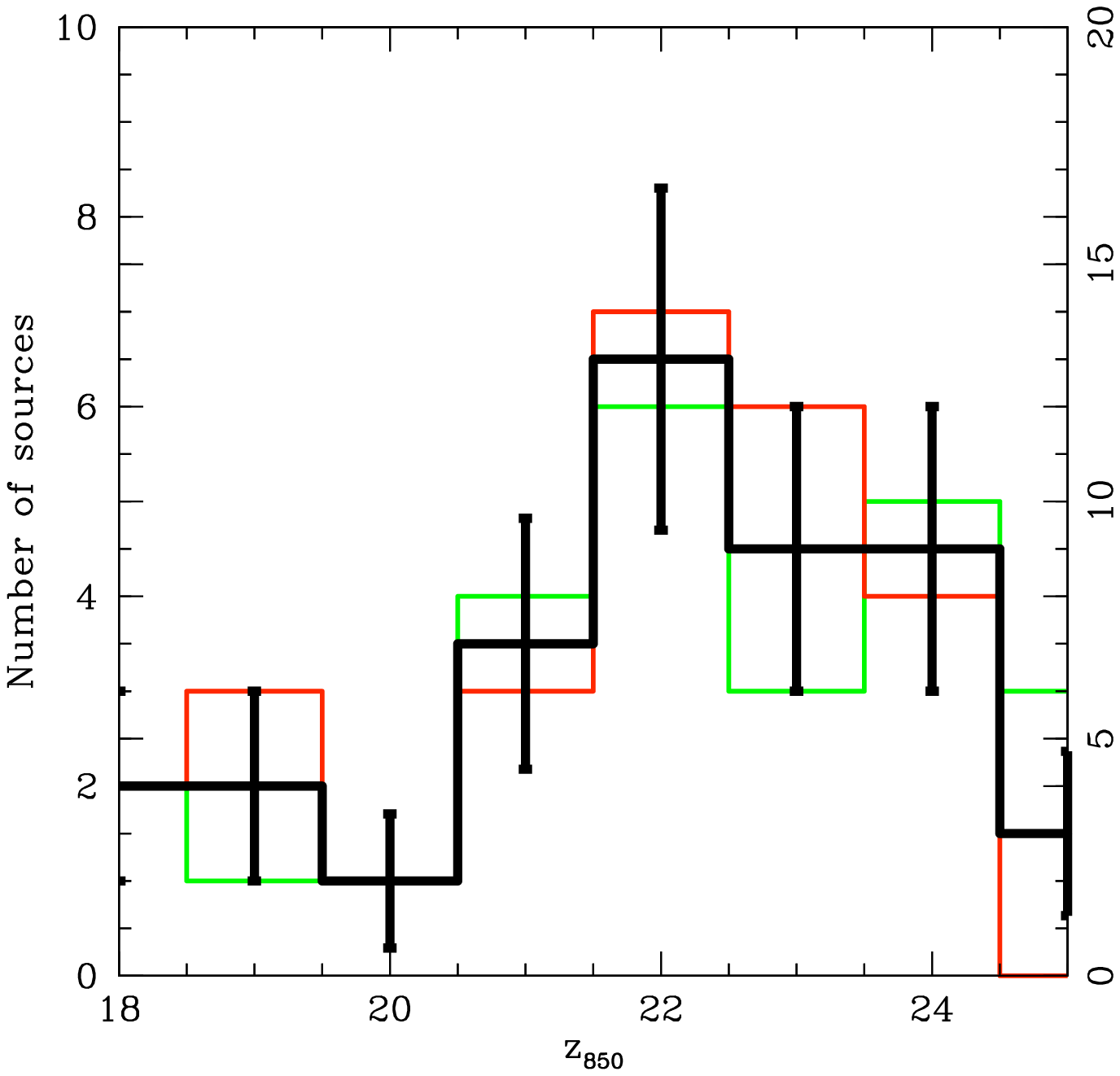}
\caption{\label{m4_star_hist_z} Histogram of the distribution of M4 and later dwarfs with $\mz < 25$ in the PEARS fields as a function of \mz\ magnitude. PEARS-N is shown in green, PEARS-S in red, with number of sources on the leftmost y-axis. The number of M4 and later dwarfs for both PEARS fields is shown plotted in black with the number of sources shown on the rightmost y-axis.}
\end{figure}

Concentrating on the lower mass stars that we spectroscopically identified  (down to the brighter limit of  $\mz < 25$) in our sample, Tables \ref{tableN} and \ref{tableS} list the stars with spectral types ranging from M0 to M9. The spectral types listed in these tables are averages of the spectral types obtained from different observations. Restricting ourselves to the latest-type objects in this group, with stellar type between M4 and M9, we spectroscopically identified 21 and 22 stellar dwarfs with $\mz < 25.0$ in PEARS-N and PEARS-S, respectively.  Figure \ref{m4_star_hist_z} shows the distribution of these objects as a function of \mz\ magnitude. Note that since we required more than one PEARS observations for these objects, the resulting effective area is slightly reduced from the one listed in Section \ref{description}, with  areas of  41.61 ${\rm arcmin^2\ }$ and 59.50  ${\rm arcmin^2\ }$, for the northern and sourthern fields,  respectively. We therefore identified 0.50 and 0.37 star per ${\rm arcmin^2\ }$ in the north and south fields respectively, and 24\% more of these dwarfs in the northern field than in the southern field. 
Examining the number of M0 to M9 dwarfs, with 51 and 63 objects in the north and south fields, the equivalent numbers are 1.23 and 1.06 star per   ${\rm arcmin^2\ }$, with a slightly lower excess of dwarfs in the northern field (11\%). The higher number of such stars in the PEARS-N fields is somewhat puzzling if one assumes that the Sun lies 6-30 pc above the Galactic Plane \citep{joshi2007,chen2001, binney1997}, which would lead one to expect more stars to be seen in the southern field. The excess of old stars in GOODS-N versus GOODS-S was previously observed by \citet{stanway2008} who found that they observed 30\% more stellar sources in the GOODS-N field. Our spectroscopically confirmed sample, using deeper broad band imaging of these fields, and which is free of extra Galactic interlopers at fainter magnitudes, therefore confirms this excess of stars in the northern field. However, in our case, the difference in observed counts between the PEARS-N and the PEARS-S field is within the uncertainties expected from simple Poisson statistics and might simply be the result of the somewhat limited surface areas covered by the PEARS survey. The likely cause of this excess is addressed further in Section \ref{dwarfs}.

\begin{figure}[h]
\includegraphics[width=7.0in]{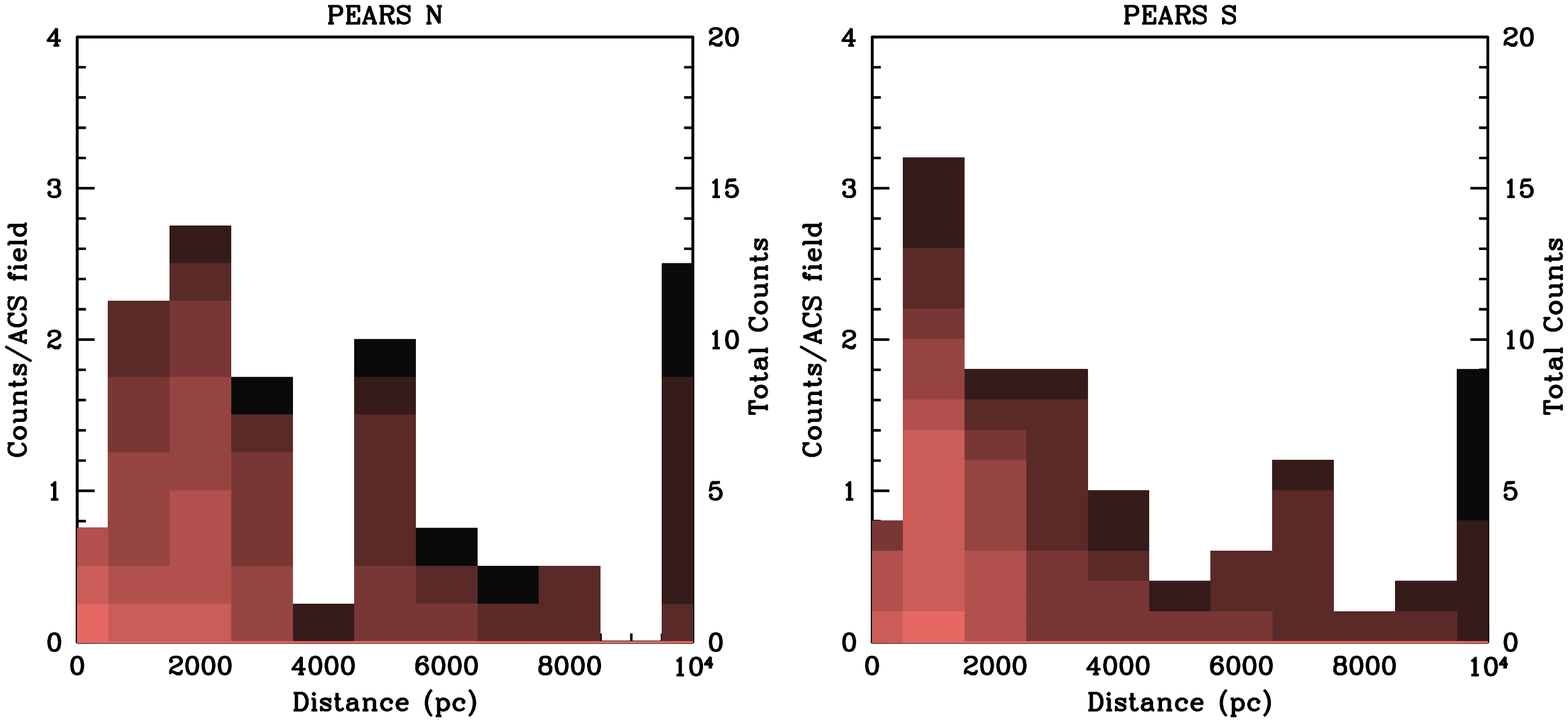}
\caption{\label{DistancesNS} Histograms of the M0 and later stars in PEARS-N (left) and PEARS-S (right). Eight histograms are overplotted using increasingly darker shades and corrresponding to the distribution of M0 and later stars with \mz\ brighter than 18, 19, 20, 21, 22, 23, 24, and 25, respectively. }
\end{figure}

Figure \ref{DistancesNS} shows the distributions of M dwarfs as a function of magnitude. The distribution of these object is seen to peak at $\mz \approx 22$. But as we discussed earlier, we are able to distinguish between stars and extra Galactic sources both morphologically (Figure \ref{R50}) and spectroscopically (Figure \ref{fracAll01})  down to much fainter magnitudes ($\mz \approx 25.0$) and lower \netsig values.  The observed decrease from $12 \pm 3.5$ at \mz=22 down to $3 \pm 1.7$, at \mz=25\ , where the errors reflect simple count statistics is therefore statistically significant and must be a reflection of the structure of our galaxy. The next section explores this is more detail.

\subsection{Number of M dwarfs and the disk scale height}\label{dwarfs}
The increased area covered by the PEARS fields over that of the much smaller HUDF/GRAPES fields discussed in  Paper I,  allows us to re-examine the Galactic thin disk scale height estimate, using a significantly larger area (8x), and two uncorrelated line of sights.  In the following two sub-sections, we first examine the distribution of the lower mass M4 to M9 dwarfs and compare it the previous work. We then extend this study to include brighter and more massive dwarfs by including M0 to M9 dwarfs.
All of the models discussed below assume that the Sun is located 8500 pc from the Galactic Center and lies 27 pc above the Galactic plane \citep{chen2001}. We used a \mz\ band luminosity function from \citet{bochanski2008} with $6.5 < M_z < 12.5$ and a J band luminosity function from \citet{cruz2007} with $10.75 < M_J < 13.75$ in the near-infrared, converting the
latter to a z-band luminosity function using the mean magnitudes and
colors of M dwarfs as a function of spectral type from  \citet{west2004}. Extinction was included, using the HI and H2 Galactic models of \cite{amores2005} and by computing ${\rm A_z}$ following \cite{Schlegel1998}, but the amount of extinction in these fields is very small ($E(B-V)=0.2-0.4$) and does not affect the fits significantly.

\subsubsection{M4-M9 dwarfs}

\begin{figure}[h]
\includegraphics[width=5.0in]{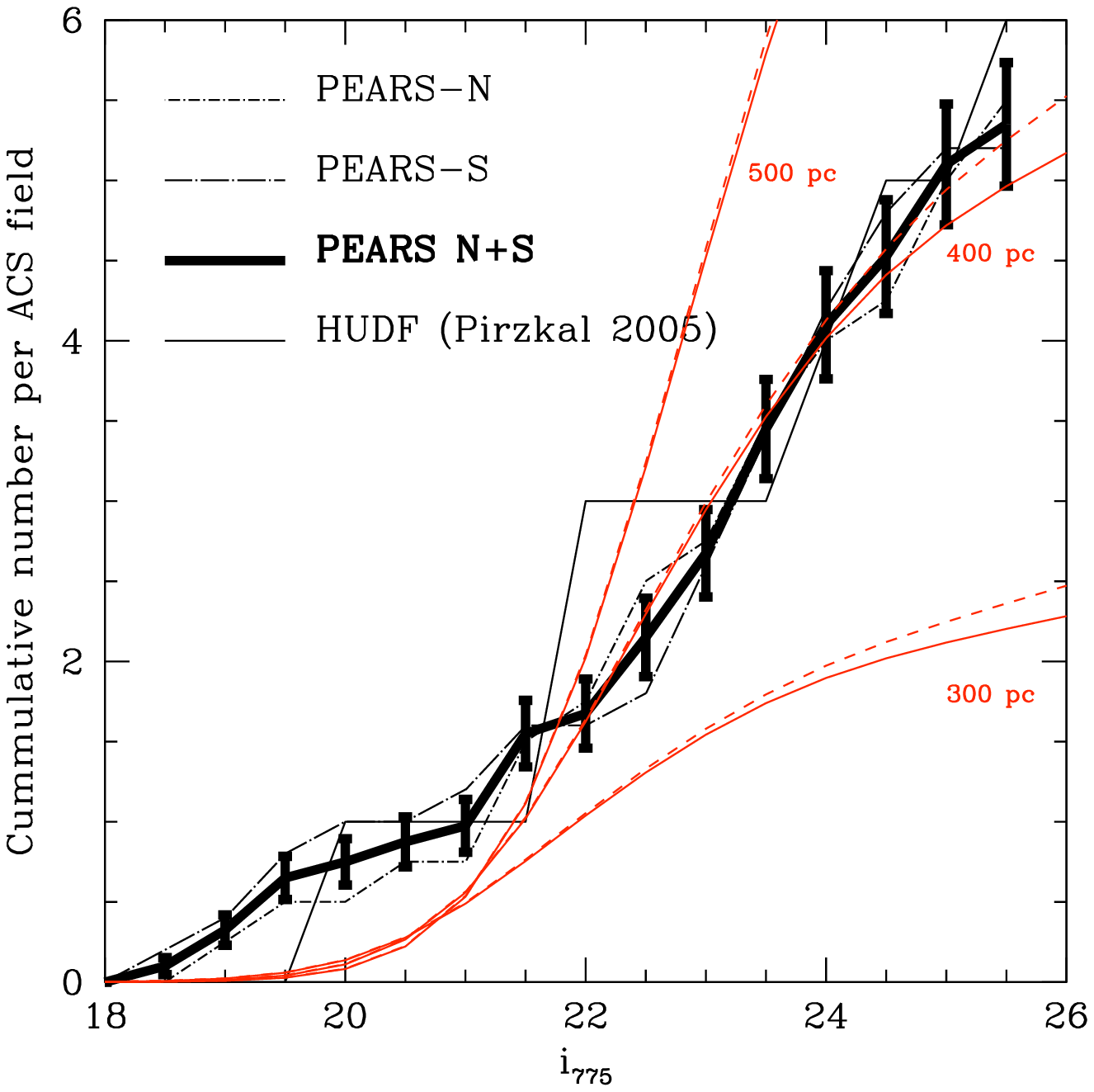}
\caption{\label{M4cumhist} The cumulative number of stars with stellar type M4 or later as a function of magnitude, normalized to the area of the HUDF. PEARS-N and PEARS-S are  shown in dash lines. The solid line shows the result obtained from the deeper HUDF data from Paper I. The actual observed number of sources in each field are the number shown per ACS field time 1, 4 and 5 for the HUDF, PEARS-N and PEARS-S, respectively. Expected cumulative  number of stars, as computed in Paper I, are shown. The PEARS data are in good agreement with the earlier GRAPES observations and exclude Galactic thin disk scale heights of ${\rm h_{thin}= 200,300, and\ 500 pc}$.}
\end{figure}

\begin{figure}[h]
\includegraphics[width=5.0in]{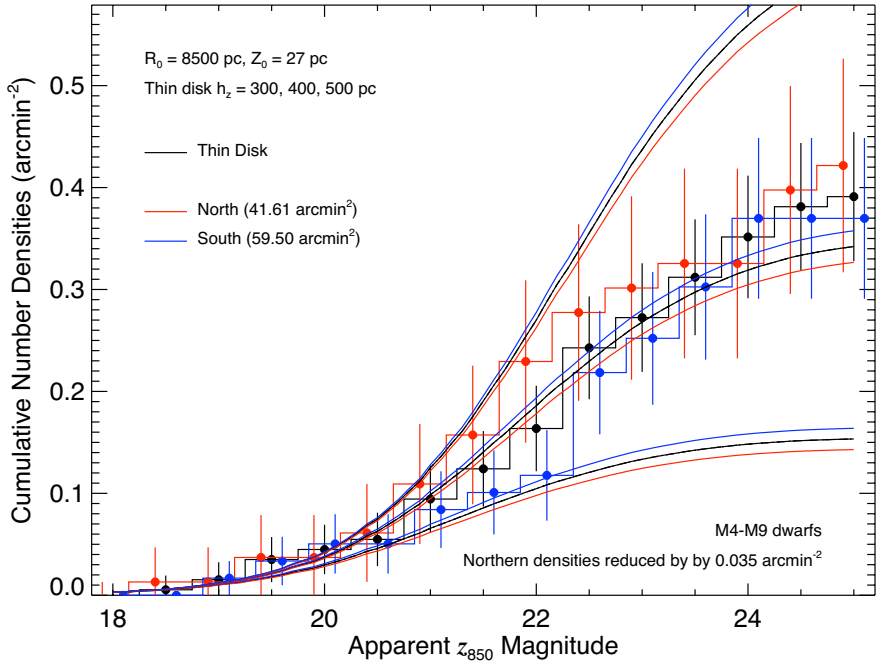}
\caption{\label{m4m9thinshift}The cumulative number of stars with stellar type M4 or later as a function of magnitude, after correcting the PEARS-N distribution for an apparent over abundance of ${\rm 0.035\ arcmin^{-2}}$ sources. The models shown are for thin disk distributions for the PEARS-N (red), PEARS-S (blue) and combined PEARS N+S (black) fields. The observations are best fitted by a disk scale height of ${\rm h_{thin} = 420 \pm 20 pc}$.}
\end{figure}

\begin{figure}[h]
\includegraphics[width=5.0in]{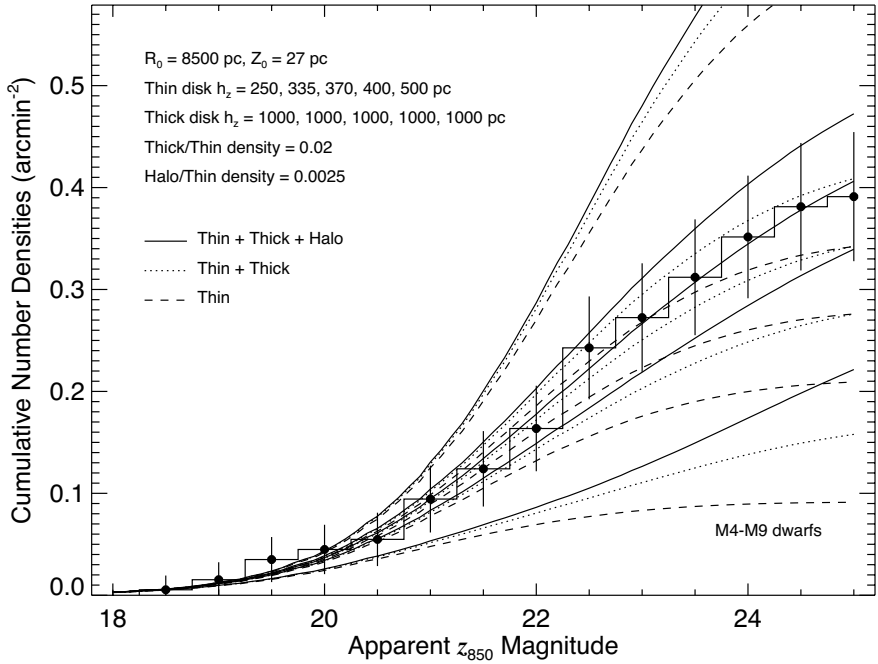}
\caption{\label{m4m9thinthickhalototalshift}The cumulative number of stars with stellar type M4 or later and models including the contribution of a Galactic thin disk, thick disk, and halo. A relative stellar density relative to the thin disk of 0.02 and 0.0025 are assumed for the thick and halo components, respectively. The slight inflection in the number of sources with $\mz > 22$ is slightly better fitted than one considering the simpler model shown in Figure \ref{m4m9thinshift}. In this case, the best model yields a thin disk scale height of ${\rm h_{thin} = 370 \pm 35 pc}$}
\end{figure}

Figure \ref{M4cumhist} shows the distribution of the lower mass M dwarfs in both the PEARS-N and PEARS-S, as well as the previous results based on GRAPES/HUDF data alone from Paper I. As shown in this figure, the PEARS results are consistent, both when comparing the northern and southern fields to each other or to the HUDF observations. The data are reasonably well fitted by a thin disk with a scale height of ${\rm \approx 400 pc}$. We however observe a slightly larger number of sources in the northern field than in the southern field.  As observed, it is difficult to fit the observed shape of the PEARS-N distribution using a smooth axisymmetric disk model. We observe a density difference between the PEARS-N and PEARS-S fields of about 0.035 ${\rm arcmin^{-2}}$, or about 1.5 objects. We do in fact observe this number of faint dwarfs at distance closer than 500 pc in the PEARS-N field. This is to be somewhat expected since the PEARS survey samples a rather small area of the sky. At distances of about 500pc, the projected area amounts to about ${\rm 1 pc^2}$ while the mean stellar spacing in the Solar neighborhood is on the order of about ${\rm 1 pc}$.  Our smooth models of the galactic structure predict a very small number of stars at these short distances and when one star is observed this significantly increases the cumulative distribution throughout. In order to compensate for this sampling problem, we simply subtracted these 1.5 extra sources from the northern cumulative distribution. This brought both the the PEARS-N and PEARS-S into good agreement at all magnitudes leaving us with no hint of different Galactic structures between the PEARS-N and PEARS-S fields, as shown in Figure \ref{m4m9thinshift}.  We determined that the best fit  thin disk model is a disk with a scale height of ${h_{thin} = 420 \pm 20 {\rm pc}}$. The inflection of the number counts around \mz=22 is however slightly better fitted if we include a combination of thin, thick and halo components to the model, as shown in Figure \ref{m4m9thinthickhalototalshift}. The data are well fitted by a thin disk with a scale height of ${\rm h_{thin} = 370 \pm 35 pc}$, together with an additional thick disk component with  ${h_{thick} = 1000 {\rm pc}}$, as well as a halo component, with assumed relative stellar density (relative to the thin disk density) of 0.02 and 0.0025, respectively. We note than an intermediate model containing only a thin and a thick disk component lead to an estimated thin disk scale height of ${\rm 390 \pm 30 pc}$. The observed trend is therefore that failure to include either a thick disk or halo component tend to lead to slightly higher scale height estimates for the thin disk, mainly because of the large number of faint sources present that simply cannot be accounted for using a simple thin disk model. These scale heights estimates are however dependent on our assumptions of specific relative stellar densities between the thin disk, the thick disk, and the halo. Making other assumptions about the relative stellar densities does lead to slightly different values for the think disk scale height. We therefore compared our observations to models using a wider range of possible halo to thin disk densities (${\rm f_{halo}=0., 0.00025, 0.0005, 0.00075}$), thick to thin disk densities (${\rm f_{thick}=0.02,0.05, 0.08}$), thick disk scale heights (${\rm h_{thick}=500, 1000, 1500, 2000 pc}$), while allowing the scale height of the thin disk to vary between 0 and 500 pc. We found all models with values of ${\rm f_{halo}=0.00075}$, or ${\rm f_{thick}=0.08}$, or ${\rm h_{thick}=2000 pc}$ to be excluded (based on a ${\rm \chi^2\ p\ value<0.05}$), leading to poor fits to the observations at the faint end ($\mz > 22$). While a range of  models are seen to be able to properly fit the bright end of the observed distribution ($\mz<22$),  many of the models with extreme values of ${\rm h_{thick},\ f_{thick}\ and\ f_{halo}}$ very poorly reproduce the observed distribution on the faint end ($\mz > 22$).  Best fits where obtained with ${\rm f_{halo}=0.00025}$, ${\rm f_{thick}=0.02}$, and ${\rm h_{thick}=1000 pc}$. Varying the values of ${\rm f_{halo}, f_{thick},\ and\ h_{thick}}$ led to a variation of the derived values of ${\rm h_{thin}}$ by only ${\rm \pm 30\ pc}$, each. This estimate of ${\rm h_{thin}}$  is robust and quite independent to our estimates of the other model parameters, in the sense that even for extreme values of  ${\rm h_{thick},\ f_{thick}\ and\ f_{halo}}$ models that could not be rejected statistically, as described above, all had rather similar values of   ${\rm h_{thin}}$. We found that the value of the scale height of the thin disk, including random and systematic errors, is ${\rm h_{thin}=370^{+60}_{-65}\ pc}$.

\subsubsection{M0-M9 dwarfs}
\begin{figure}[h]
\includegraphics[width=5.0in]{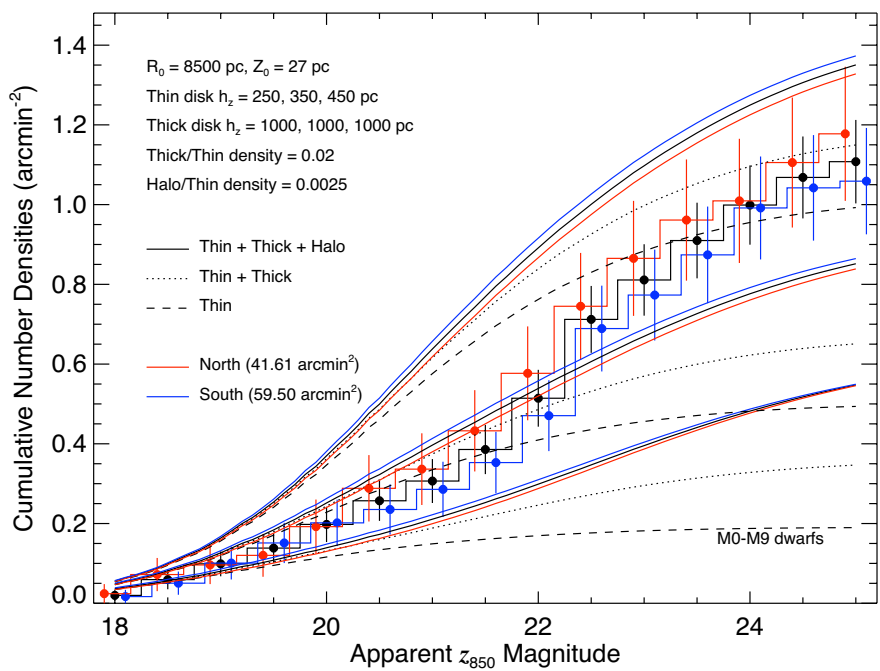}
\caption{\label{m0m9thinthickhalo}This figure shows the cummulative distribution of all M dwarf stars. Unlike Figures \ref{m4m9thinshift} and \ref{m4m9thinthickhalototalshift}, the observed cumulative distributions are not well fitted by our simple modes. The number of faint, and more distant sources is significantly higher than what would be expected using the thin, thick and halo parameters determined using the lower mass M4 and later population of stars.}
\end{figure}

\begin{figure}[h]
\includegraphics[width=5.0in]{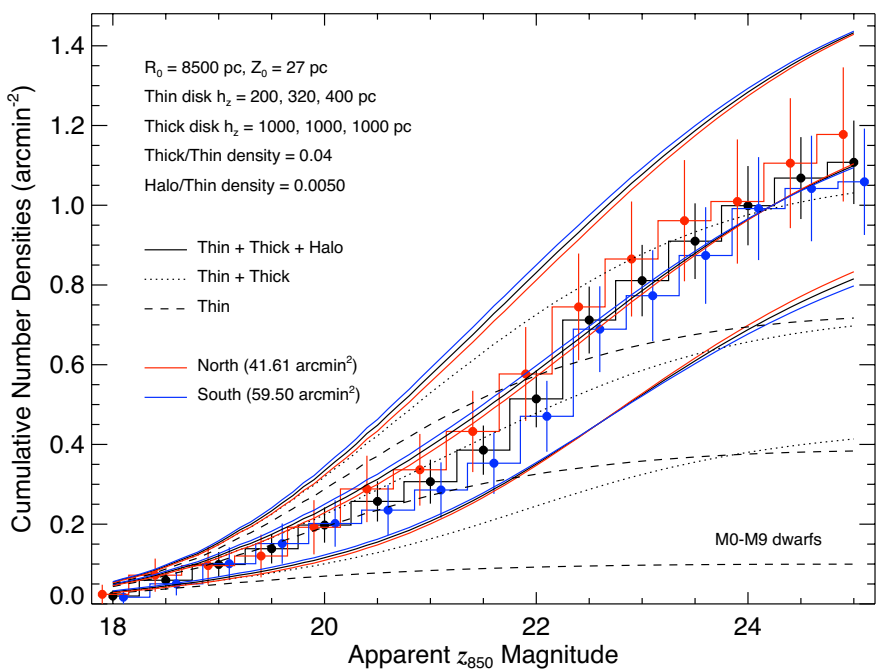}
\caption{\label{m0m9thinthickhalohigh}This Figure shows the effect of considering denser thick and halo components. As show here, doubling the stellar densities of these component, compared to what we used in Figures \ref{m4m9thinthickhalototalshift} and \ref{m0m9thinthickhalo} better reproduce our observations with a thin disk scale height slightly reduced to ${\rm h_{thin} = 320}$.}
\end{figure}

\begin{figure}[h]
\includegraphics[width=5.0in]{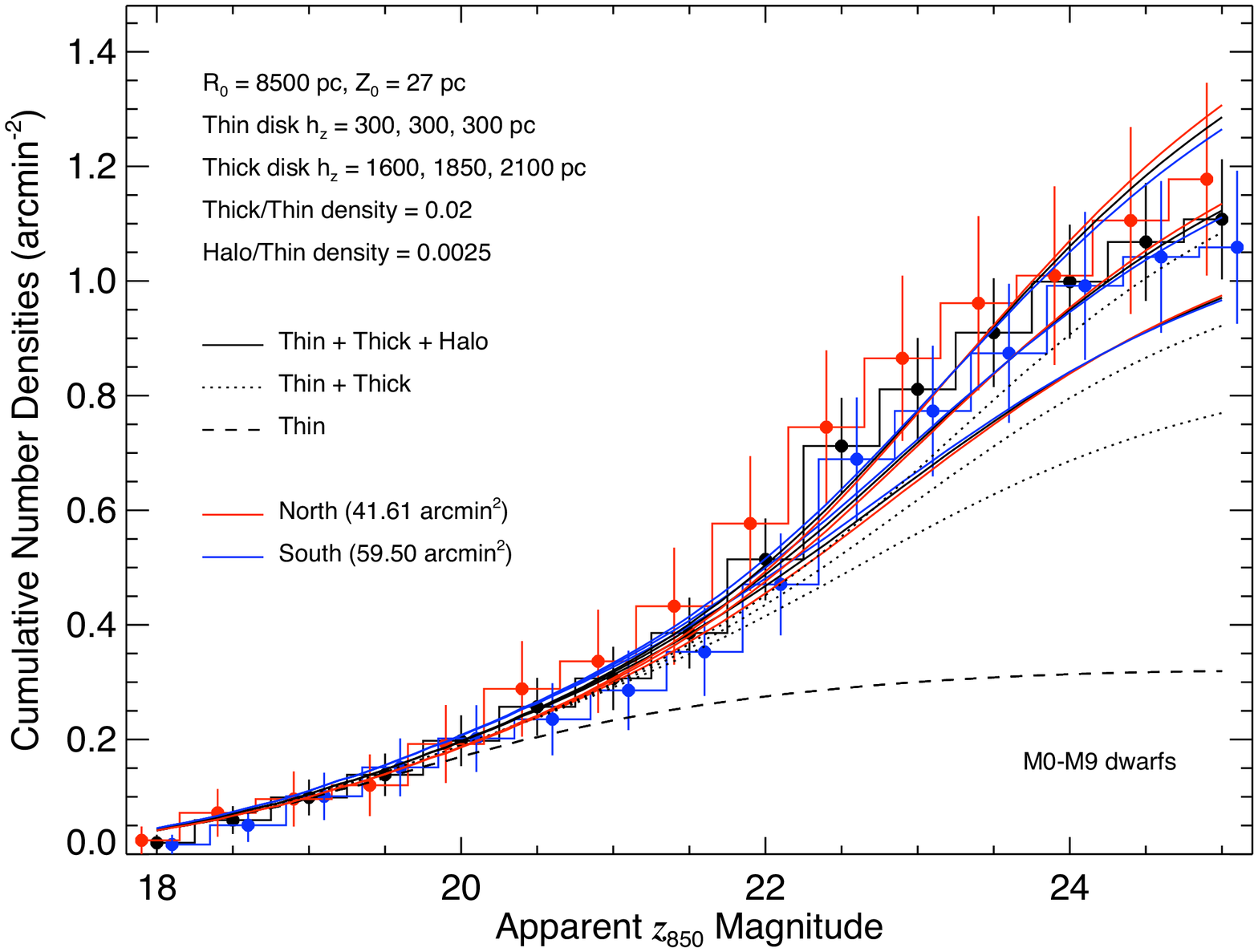}
\caption{\label{m0m9thinthickhalotdthick}While Figure \ref{m0m9thinthickhalohigh} shows that a fit to the observations can be obtained by considering denser thick and halo component, this Figure illustrates how one can instead choose to increase the scale height of the thick disk component. Indeed, keeping the nominal relative densities for the thick disk and halo components down to 0.02 and 0.0025 (relative to the thin disk), an increase of the thick disk scale height of ${\rm h_{thick} = 1850 \pm 250 pc}$ coupled with a thin disk scale height of ${\rm h_{thin} = 300 pc}$ is a good fit to the data.}
\end{figure}

The inclusion of more massive, brighter M dwarfs in our sample, allows us to probe to significantly larger distances (${\rm < 10 kpc}$), where one would expect the contribution of either (or both) a thick disk or a halo to be increasingly significant. As shown in Figure \ref{m0m9thinthickhalo}, the inclusion of more massive dwarfs results in an observed distribution containing a larger number of fainter sources, causing a significant inflection in  the distribution of these sources around $\mz=22$. As shown in this Figure, the data are not so well fitted by the models that we previously established using the M4 to M9 population of stars. In fact, we found that one way to obtain a better fit to the observations was by allowing the relative densities of the thick disk and of the halo to increase by a factor of 2. This is shown in Figure \ref{m0m9thinthickhalohigh}. Assuming a denser thick disk and halo, the inferred thin disk scale height is reduced to ${\rm h_{thin} = 320 pc}$. The higher than nominal thick disk and halo densities could be an indication that the halo density for cool dwarfs is higher than what is commonly derived for more massive stars. While this might not be an unreasonable conclusion since the halo is more than 12 Gyr old, and could be hence expected to contain more low mass stars than massive stars, we note that we were able to  equally well fit the observations using a different model. As we show in Figure \ref{m0m9thinthickhalotdthick}, we could keep the relative densities of the thick disk and halo unchanged and choose  instead to  increase the scale height of the thick disk component from ${\rm h_{thick} = 1000 pc}$ to ${\rm h_{thick} = 1850 \pm 250 pc}$. In the latter case, the preferred thin disk scale height becomes ${\rm h_{thin} = 300 pc}$. In fact, we found that one can carefully chose a set of densities and scale heights in order to best fit the data, but there are a strong degeneracies between these parameters. One can however observe that our attempts to fit the M0 to M9 dwarf distribution systematically lead us to derive thin disk scale heights that are smaller than what we derived using only M4 to M9 dwarfs. While this could be tentative evidence for an increase in the scale height of the thin disk for lower mass stars, the significance of this difference in scale height is small once we account  for systematic errors caused by specific choices of the relative stellar densities between halo, thin, and thick disk. Proceeding as we did in the previous section (rejecting models with an ${\rm \chi^2\ p\ value<0.05}$), and assuming a range of possible values for ${\rm f_{halo}, f_{thick},\ and\ h_{thick}}$, we derive a best estimate of ${\rm h_{thin} = 300^{+70}_{-70}\ pc}$. Best fits were obtained with values of ${\rm f_{halo}=0.0005}$, ${\rm f_{thick}=0.05}$, and ${\rm h_{thick}=1000 pc}$. We are able to exclude values such as ${\rm f_{halo} \le 0.00025, f_{thick} \ge 0.08,\ and\ h_{thick}\neq 1000\ pc}$ as these furthermore do not properly fit the shape of the fainter end ($\mz > 22$) of the observed distribution of M0-M9 dwarfs. We also note that our results are consistent to the results from \citet{juric2008}, using a significantly smaller field of view but down to significantly deeper limiting magnitudes. Probing to very faint magnitudes, the scale height of the thin disk, as traced by cool M0 to M9 dwarfs, is confirmed to be in the range of  ${\rm h_{thin} \approx 300-400 pc}$.

\subsection{Non M dwarf objects}
As shown in Figure \ref{star_select_hist}, the vast majority of the stars identified in the PEARS regions are M dwarf stars. The spectral identification of earlier type stars and with a stellar type later than B is not accurate enough to identify the spectral type of these stars, especially at faint magnitudes and low values of \netsig. Dwarf stars of spectral type late than  M are are equally difficult to identify because these objects are expected to be very faint and remain nearly undetected except in the reddest bands (\mz). We do identify two faint, red stars that appear to be best fitted by L dwarf templates in the PEARS-N field. These two objects (ID 42016 and 75398) have magnitudes of \mz=24.49 and \mz=24.20, are 372 and 281 pc away, and have estimated stellar types of L6 and L7 respectively. We also identify a slightly fainter L2 dwarf at \mz=25.1 (ID 89268) at an estimated distance of 1100pc.  We detect no L and later dwarf brighter than \mz=25 in the PEARS-S field, and only marginally identify two sources (ID 48362 and 67495) with \mz=25.7 and 25.6, with spectral types of L2 and L4, and at distances of 1500 and 900pc, respectively.  Down to \mz=25, the detection of two L dwarfs in the PEARS fields is however consistent with the numbers that would be expected from the work of \citet{caballero2008}.  Once adjusting for the smaller field of view, the latter would expect  $\approx 1.5$\ L0 and later dwarfs with $\mz<25$ in the combined PEARS areas. The low number of identified L dwarf and later type stars does not allow us to constraint the properties of these at low stellar masses. Identifying L and T dwarfs using ACS data remains difficult because of the rather extreme red colors of these sources. Future instruments on board of HST, such as the Wide Field Camera 3 (WFPC3) with its infrared bands and slitless spectroscopic mode, should greatly facilitate the studies of such sources. Parallel observations using WFC3 in particular might identify a significant number of these sources in the future.

While the goal of this paper is not to seriously contemplate the population of White Dwarfs in these fields, and their possible contribution to a dark matter Galactic Halo, as we did in Paper I, we can nevertheless examine whether any reasonable constraints can be set. Indeed, we can isolate a population of stars as being potentially White Dwarfs by restricting ourselves to stars that lie in the appropriate color-color region of  Figure \ref{izvi} (i.e. $\mi-\mz < 0.15$\ and $\mv-\mi < 1$). We identify 23 and 25 objects in the PEARS-N and PEARS-S, respectively and with $\mz < 25$.  For comparison, the total number of stars down to that magnitude limit is 95 and 108 in PEARS-N and PEARS-S, respectively. Down to $\mz=25$, White Dwarfs should remain detectable up to a maximum distance of  $\approx {\rm 400pc}$. Within the solid angle of the combined PEARS fields, and 400pc, we would expect this observable volume to contain  a mass of $1.7 M_\sun$ of whatever the Galactic dark matter Halo might be made up of  \citep{kawaler1996}. If this Halo was 100\% of 0.6 $M_\sun$ White Dwarfs, we would therefore expect at most about 3 White Dwarfs. Moreover, and as we determined in Paper I, we should not expect more than 10\% of the mass of the dark matter Halo to be in the form of White Dwarfs, so that the expected number should reasonably be expected to be much less than one. The PEARS data simply do not allow us to probe the stellar population down to faint enough magnitudes  in turn allowing us to probe a significantly part of the Halo. We may be able to further exclude some of the PEARS non-M Dwarfs however. For example, we can exclude stars that are well fitted by  main sequence stellar spectra and whose photometric parallax places well within a thin or thick disk component (e.g $<20000$pc). Still, this only allows us to decrease the number of White Dwarf candidates down to 10 in the PEARS-N field and 11 in the PEARS-S field, which is, again, more than required to populate the dark matter halo. What is required is the ability to securely place these objects within the Halo via the unambiguous detection of the expected proper motion of these objects if they were within the Halo.  The GOODS 2.0 data however were not assembled in a way that allows us to measure proper motion by simply comparing GOOODS 1.0 and GOODS 2.0 images. The GOODS 2.0 documentation warns user that stars with large proper motion might be missing from the final GOODS 2.0 images due to the manner in which new imaging data were simply combined with GOODS 1.0 data. With a magnitude limit of \mz=25, and a maximum distance of 400pc, we would  expect any Halo object (with expected tangential velocities of several 100 km/s) to move significantly between the first and last epoch of the GOODS observations. We checked that the PEARS stellar candidates were present in both the GOODS 2.0 and GOODS 1.0 images and that their flux, and morphology was not significantly different. A more careful analysis of the proper motion of these sources is warranted to close this discussion and this will be possible if and when the GOODS data are combined together to form a time sequence of \mi\ and \mz\ band images.

\section{Conclusion}
By combining the new and deeper imaging of the GOODS-N and GOODS-S fields
(GOODS 2.0) with new deep ACS slitless spectroscopy, we have selected a robust sample
of old low mass stars using a self-consistent set of spectroscopic and morphological
selection criteria. By combining spectroscopy and morphological
measurements, we were able to refine the morphological selection of objects in these
fields down to $\mz = 26$ mag. It was also demonstrated that stars can be
spectroscopically  identified to the faint magnitude of $\mz = 25$ by fitting stellar
templates to the slitless spectra.
Through the identification of a sample of cool low mass dwarfs in these fields that is
free of faint extragalactic interlopers, we estimate that the scale height of the
Galactic thin disk, as traced by M4-M9 dwarfs, is ${\rm h_{thin} = 370^{+60}_{-65}\ pc}$. When including
the slightly more massive M0 to M4 dwarfs in our sample, we derive ${\rm h_{thin} = 300 \pm 70 pc}$. We also showed that a simple
thin disk model leads to an under-prediction of the number of such faint sources. A combination of thick disk and halo components is required to reproduce the observed counts 
of M0 to M9 dwarfs, even though the current observations do not set new strong constraints of the scale height and relative densities of these components.

\begin{deluxetable}{cccccccc}
\tabletypesize{\footnotesize} 
\tablecaption{M0 and later stars in PEARS-N with $\mz < 25$. \label{tableN}}
\tablehead{\colhead{PID} & \colhead{RA (J2000)} & \colhead{Dec (J2000)} & \colhead{\mb} & \colhead{\mb-\mv}   &  \colhead{\mv-\mi} & \colhead{\mi-\mz} & \colhead{Spectral Type}
}
\startdata  
n40790  &  189.21332946  &  62.1869242  &  $23.28 \pm 0.004$  &  $0.73 \pm 0.046$  &  $1.69 \pm 0.019$  &  $0.69  \pm 0.005$  &  M4.0  \\
n41952  &  189.21657649  &  62.188669742  &  $23.26 \pm 0.003$  &  $2.12 \pm 0.154$  &  $1.93 \pm 0.018$  &  $0.68  \pm 0.004$  &  M6.0  \\
n42294  &  189.18393871  &  62.189816961  &  $21.56 \pm 0.001$  &  $2.25 \pm 0.046$  &  $1.77 \pm 0.006$  &  $0.63  \pm 0.002$  &  M4.0  \\
n43597  &  189.21252896  &  62.192935832  &  $21.33 \pm 0.001$  &  $2.09 \pm 0.016$  &  $1.49 \pm 0.002$  &  $0.47  \pm 0.001$  &  M3.0  \\
n44285  &  189.25323563  &  62.193551908  &  $23.39 \pm 0.004$  &  $2.59 \pm 0.169$  &  $1.61 \pm 0.014$  &  $0.49  \pm 0.005$  &  M4.0  \\
n45100  &  189.23315668  &  62.194724655  &  $22.27 \pm 0.002$  &  $1.72 \pm 0.018$  &  $0.91 \pm 0.003$  &  $0.23  \pm 0.003$  &  M0.0  \\
n50115  &  189.25641523  &  62.203260841  &  $24.70 \pm 0.009$  &  $71.60 \pm 99.000$  &  $2.09 \pm 0.059$  &  $0.62  \pm 0.013$  &  M5.0  \\
n50444  &  189.23072083  &  62.203807475  &  $24.20 \pm 0.006$  &  $3.62 \pm 0.854$  &  $1.84 \pm 0.031$  &  $0.65  \pm 0.008$  &  M5.5  \\
n53300  &  189.1703926  &  62.209518637  &  $23.87 \pm 0.006$  &  $2.29 \pm 0.115$  &  $1.29 \pm 0.019$  &  $0.31  \pm 0.009$  &  M2.0  \\
n53956  &  189.19487478  &  62.210392327  &  $22.03 \pm 0.002$  &  $1.94 \pm 0.032$  &  $1.35 \pm 0.005$  &  $0.45  \pm 0.002$  &  M2.3  \\
n54158  &  189.16758171  &  62.212045805  &  $20.67 \pm 0.001$  &  $2.14 \pm 0.025$  &  $1.75 \pm 0.003$  &  $0.62  \pm 0.001$  &  M4.0  \\
n55079  &  189.22517869  &  62.212753439  &  $22.36 \pm 0.002$  &  $1.70 \pm 0.037$  &  $1.47 \pm 0.006$  &  $0.48  \pm 0.003$  &  M3.0  \\
n54828  &  189.20132337  &  62.213929254  &  $20.43 \pm 0.001$  &  $2.02 \pm 0.013$  &  $1.71 \pm 0.002$  &  $0.62  \pm 0.001$  &  M4.0  \\
n59383  &  189.17548928  &  62.219577156  &  $23.40 \pm 0.004$  &  $2.09 \pm 0.071$  &  $1.31 \pm 0.011$  &  $0.38  \pm 0.006$  &  M1.3  \\
n60144  &  189.2280103  &  62.224514106  &  $18.92 \pm 0.000$  &  $1.70 \pm 0.002$  &  $0.89 \pm 0.000$  &  $0.28  \pm 0.000$  &  M0.5  \\
n69781  &  189.2410194  &  62.23840465  &  $21.82 \pm 0.001$  &  $2.05 \pm 0.021$  &  $1.12 \pm 0.003$  &  $0.40  \pm 0.002$  &  M2.0  \\
n69090  &  189.19545289  &  62.239549034  &  $19.68 \pm 0.000$  &  $1.69 \pm 0.003$  &  $0.93 \pm 0.001$  &  $0.30  \pm 0.000$  &  M1.0  \\
n71802  &  189.13038078  &  62.241443847  &  $24.13 \pm 0.006$  &  $1.82 \pm 0.080$  &  $1.22 \pm 0.014$  &  $0.31  \pm 0.009$  &  M2.0  \\
n72344  &  189.14806812  &  62.242105605  &  $24.79 \pm 0.009$  &  $2.94 \pm 0.338$  &  $1.31 \pm 0.024$  &  $0.39  \pm 0.014$  &  M3.0  \\
n72236  &  189.14399543  &  62.242446538  &  $22.98 \pm 0.003$  &  $1.76 \pm 0.021$  &  $0.97 \pm 0.005$  &  $0.27  \pm 0.004$  &  M1.0  \\
n74900  &  189.15675651  &  62.248291218  &  $21.99 \pm 0.002$  &  $1.52 \pm 0.103$  &  $2.45 \pm 0.020$  &  $1.02  \pm 0.004$  &  M7.0  \\
n75295  &  189.18382862  &  62.249931788  &  $20.62 \pm 0.001$  &  $2.33 \pm 0.021$  &  $1.80 \pm 0.002$  &  $0.62  \pm 0.001$  &  M4.0  \\
n76018  &  189.17091061  &  62.250340871  &  $21.90 \pm 0.001$  &  $1.94 \pm 0.025$  &  $1.54 \pm 0.005$  &  $0.54  \pm 0.002$  &  M4.0  \\
n76445  &  189.12650964  &  62.25074679  &  $21.50 \pm 0.001$  &  $2.19 \pm 0.034$  &  $1.92 \pm 0.005$  &  $0.73  \pm 0.002$  &  M6.0  \\
n78157  &  189.22998288  &  62.253572335  &  $22.96 \pm 0.003$  &  $2.03 \pm 0.074$  &  $1.50 \pm 0.009$  &  $0.48  \pm 0.005$  &  M3.0  \\
n79055  &  189.12913756  &  62.257813653  &  $20.21 \pm 0.001$  &  $1.82 \pm 0.008$  &  $1.35 \pm 0.001$  &  $0.48  \pm 0.001$  &  M2.7  \\
n82422  &  189.22754323  &  62.260185539  &  $24.86 \pm 0.009$  &  $1.89 \pm 0.163$  &  $1.27 \pm 0.024$  &  $0.32  \pm 0.014$  &  M2.3  \\
n83506  &  189.32600578  &  62.262931747  &  $23.62 \pm 0.004$  &  $1.85 \pm 0.055$  &  $1.10 \pm 0.009$  &  $0.27  \pm 0.006$  &  M1.0  \\
n85549  &  189.23085947  &  62.26642455  &  $22.76 \pm 0.002$  &  $1.71 \pm 0.022$  &  $0.89 \pm 0.004$  &  $0.25  \pm 0.003$  &  M0.0  \\
n86491  &  189.28614303  &  62.269962435  &  $20.29 \pm 0.000$  &  $1.71 \pm 0.004$  &  $0.92 \pm 0.001$  &  $0.30  \pm 0.001$  &  M1.0  \\
n87140  &  189.34336558  &  62.270233245  &  $21.11 \pm 0.001$  &  $1.82 \pm 0.015$  &  $1.38 \pm 0.003$  &  $0.45  \pm 0.001$  &  M3.0  \\
n89312  &  189.31862752  &  62.275029386  &  $21.36 \pm 0.001$  &  $2.04 \pm 0.034$  &  $1.80 \pm 0.005$  &  $0.62  \pm 0.001$  &  M4.0  \\
n90163  &  189.28378636  &  62.275443643  &  $21.83 \pm 0.001$  &  $1.84 \pm 0.013$  &  $0.88 \pm 0.002$  &  $0.25  \pm 0.002$  &  M1.0  \\
n90991  &  189.32232136  &  62.276580238  &  $22.11 \pm 0.002$  &  $1.51 \pm 0.026$  &  $1.38 \pm 0.006$  &  $0.52  \pm 0.003$  &  M3.0  \\
n92893  &  189.27292455  &  62.279385303  &  $24.33 \pm 0.007$  &  $2.48 \pm 0.349$  &  $1.76 \pm 0.031$  &  $0.56  \pm 0.012$  &  M4.3  \\
n98619  &  189.33633675  &  62.290648774  &  $24.34 \pm 0.007$  &  $2.60 \pm 0.438$  &  $1.84 \pm 0.039$  &  $0.55  \pm 0.012$  &  M4.3  \\
n100774  &  189.25304865  &  62.295183229  &  $21.31 \pm 0.001$  &  $2.39 \pm 0.065$  &  $2.07 \pm 0.006$  &  $0.77  \pm 0.002$  &  M6.0  \\
n102481  &  189.39807316  &  62.297657483  &  $22.56 \pm 0.002$  &  $1.85 \pm 0.025$  &  $0.97 \pm 0.005$  &  $0.27  \pm 0.003$  &  M0.5  \\
n105087  &  189.27955342  &  62.302139471  &  $22.48 \pm 0.002$  &  $1.76 \pm 0.018$  &  $0.96 \pm 0.004$  &  $0.27  \pm 0.003$  &  M0.7  \\
n105571  &  189.40321531  &  62.304649112  &  $19.37 \pm 0.000$  &  $2.14 \pm 0.012$  &  $1.83 \pm 0.002$  &  $0.71  \pm 0.001$  &  M5.0  \\
n107697  &  189.33967607  &  62.306845915  &  $22.38 \pm 0.002$  &  $1.99 \pm 0.085$  &  $1.89 \pm 0.012$  &  $0.70  \pm 0.004$  &  M4.5  \\
n104802  &  189.42909619  &  62.308475808  &  $18.33 \pm 0.000$  &  $2.15 \pm 0.006$  &  $1.89 \pm 0.001$  &  $0.68  \pm 0.000$  &  M6.0  \\
n106609  &  189.36936868  &  62.309290375  &  $18.20 \pm 0.000$  &  $2.01 \pm 0.003$  &  $1.51 \pm 0.000$  &  $0.54  \pm 0.000$  &  M3.2  \\
n112738  &  189.31591044  &  62.315748007  &  $21.76 \pm 0.001$  &  $2.14 \pm 0.022$  &  $1.21 \pm 0.003$  &  $0.36  \pm 0.002$  &  M1.5  \\
n115093  &  189.42840809  &  62.317995543  &  $19.52 \pm 0.000$  &  $1.88 \pm 0.005$  &  $1.59 \pm 0.001$  &  $0.54  \pm 0.000$  &  M0.3  \\
n115421  &  189.30862113  &  62.318195605  &  $22.98 \pm 0.003$  &  $2.79 \pm 0.432$  &  $2.49 \pm 0.047$  &  $1.12  \pm 0.007$  &  M7.0  \\
n119309  &  189.3741763  &  62.325816919  &  $19.98 \pm 0.000$  &  $1.96 \pm 0.007$  &  $1.41 \pm 0.001$  &  $0.48  \pm 0.001$  &  M2.0  \\
n119657  &  189.31551318  &  62.332519168  &  $17.78 \pm 0.000$  &  $1.90 \pm 0.002$  &  $1.33 \pm 0.000$  &  $0.81  \pm 0.000$  &  M4.0  \\
n124069  &  189.38313312  &  62.339029918  &  $22.49 \pm 0.002$  &  $2.03 \pm 0.116$  &  $2.13 \pm 0.015$  &  $0.80  \pm 0.004$  &  M6.0  \\
n124138  &  189.41178601  &  62.339994357  &  $20.32 \pm 0.000$  &  $1.89 \pm 0.005$  &  $1.20 \pm 0.001$  &  $0.38  \pm 0.001$  &  M2.0  \\
n127769  &  189.38860463  &  62.34962752  &  $23.21 \pm 0.003$  &  $2.01 \pm 0.039$  &  $1.16 \pm 0.007$  &  $0.34  \pm 0.005$  &  M1.5  \\
\enddata
\end{deluxetable}

\begin{deluxetable}{cccccccc}
\tabletypesize{\footnotesize} 
\tablecaption{M0 and later stars in PEARS-S with $\mz < 25$. \label{tableS}}
\tablehead{\colhead{PID} & \colhead{RA (J2000)} & \colhead{Dec (J2000)} & \colhead{\mz} & \colhead{\mb-\mv}   &  \colhead{\mv-\mi} & \colhead{\mi-\mz} & \colhead{Spectral Type}
}
\startdata  
s9247  &  53.188143195  &  -27.925813928  &  $18.38 \pm 0.000$  &  $1.89 \pm 0.003$  &  $1.36 \pm 0.000$  &  $0.46  \pm 0.000$  &  M2.5  \\
s12699  &  53.178470374  &  -27.921707074  &  $19.87 \pm 0.000$  &  $1.99 \pm 0.006$  &  $1.46 \pm 0.001$  &  $0.51  \pm 0.001$  &  M3.0  \\
s17867  &  53.157062298  &  -27.912900316  &  $23.70 \pm 0.004$  &  $3.86 \pm 2.141$  &  $2.46 \pm 0.056$  &  $1.00  \pm 0.010$  &  M7.0  \\
s20923  &  53.15080478  &  -27.905462445  &  $22.43 \pm 0.002$  &  $1.79 \pm 0.020$  &  $0.94 \pm 0.004$  &  $0.29  \pm 0.003$  &  M1.0  \\
s27396  &  53.198900382  &  -27.888923584  &  $19.59 \pm 0.000$  &  $1.91 \pm 0.004$  &  $1.41 \pm 0.001$  &  $0.52  \pm 0.000$  &  M3.0  \\
s32267  &  53.16809425  &  -27.880270966  &  $20.04 \pm 0.001$  &  $1.98 \pm 0.008$  &  $1.37 \pm 0.001$  &  $0.49  \pm 0.001$  &  M2.0  \\
s32132  &  53.192616336  &  -27.880286197  &  $20.60 \pm 0.001$  &  $2.14 \pm 0.021$  &  $1.82 \pm 0.003$  &  $0.72  \pm 0.001$  &  M5.3  \\
s34492  &  53.152809587  &  -27.877872787  &  $22.29 \pm 0.002$  &  $2.57 \pm 0.104$  &  $1.73 \pm 0.010$  &  $0.61  \pm 0.004$  &  M4.0  \\
s45021  &  53.170246691  &  -27.856212762  &  $21.79 \pm 0.001$  &  $1.93 \pm 0.022$  &  $1.43 \pm 0.004$  &  $0.43  \pm 0.002$  &  M2.8  \\
s45879  &  53.15845951  &  -27.854914197  &  $22.29 \pm 0.002$  &  $1.81 \pm 0.083$  &  $2.12 \pm 0.012$  &  $0.79  \pm 0.003$  &  M5.8  \\
s47103  &  53.157761709  &  -27.852749559  &  $21.79 \pm 0.001$  &  $1.94 \pm 0.016$  &  $1.13 \pm 0.003$  &  $0.35  \pm 0.002$  &  M1.0  \\
s45452  &  53.182332457  &  -27.851835647  &  $18.95 \pm 0.000$  &  $2.08 \pm 0.006$  &  $1.75 \pm 0.001$  &  $0.61  \pm 0.000$  &  M3.5  \\
s47646  &  53.14876779  &  -27.850442614  &  $20.37 \pm 0.001$  &  $1.86 \pm 0.010$  &  $1.37 \pm 0.002$  &  $0.47  \pm 0.001$  &  M2.3  \\
s46349  &  53.210299323  &  -27.850005066  &  $18.86 \pm 0.000$  &  $2.06 \pm 0.006$  &  $1.64 \pm 0.001$  &  $0.62  \pm 0.000$  &  M3.0  \\
s48173  &  53.213537665  &  -27.850926517  &  $23.15 \pm 0.004$  &  $2.07 \pm 0.126$  &  $1.81 \pm 0.019$  &  $0.58  \pm 0.006$  &  M4.0  \\
s53237  &  53.179961391  &  -27.843028718  &  $23.49 \pm 0.005$  &  $1.70 \pm 0.123$  &  $1.96 \pm 0.027$  &  $0.65  \pm 0.008$  &  M4.0  \\
s58796  &  53.151821654  &  -27.834453455  &  $23.85 \pm 0.005$  &  $1.85 \pm 0.062$  &  $1.19 \pm 0.012$  &  $0.32  \pm 0.008$  &  M1.5  \\
s58826  &  53.173279326  &  -27.834179328  &  $22.39 \pm 0.002$  &  $2.37 \pm 0.087$  &  $1.95 \pm 0.011$  &  $0.65  \pm 0.003$  &  M4.3  \\
s63079  &  53.162325116  &  -27.826953557  &  $23.47 \pm 0.005$  &  $2.02 \pm 0.050$  &  $0.97 \pm 0.009$  &  $0.29  \pm 0.007$  &  M1.0  \\
s63028  &  53.204981843  &  -27.825889543  &  $22.15 \pm 0.002$  &  $2.14 \pm 0.070$  &  $1.82 \pm 0.009$  &  $0.63  \pm 0.003$  &  M3.7  \\
s63752  &  53.172581508  &  -27.824867765  &  $21.74 \pm 0.001$  &  $2.06 \pm 0.038$  &  $1.69 \pm 0.006$  &  $0.63  \pm 0.002$  &  M3.9  \\
s63993  &  53.147900881  &  -27.823916981  &  $21.91 \pm 0.002$  &  $1.90 \pm 0.019$  &  $1.22 \pm 0.004$  &  $0.36  \pm 0.002$  &  M2.0  \\
s66572  &  53.175309643  &  -27.819896602  &  $23.58 \pm 0.004$  &  $1.02 \pm 0.344$  &  $3.13 \pm 0.118$  &  $1.22  \pm 0.011$  &  M8.4  \\
s69522  &  53.1648152  &  -27.814370229  &  $22.64 \pm 0.002$  &  $2.25 \pm 0.108$  &  $1.96 \pm 0.014$  &  $0.77  \pm 0.004$  &  M6.0  \\
s70032  &  53.146065321  &  -27.812328869  &  $20.77 \pm 0.001$  &  $2.11 \pm 0.022$  &  $1.75 \pm 0.003$  &  $0.63  \pm 0.001$  &  M3.9  \\
s74670  &  53.161613094  &  -27.802776396  &  $23.97 \pm 0.005$  &  $1.82 \pm 0.058$  &  $0.99 \pm 0.010$  &  $0.23  \pm 0.008$  &  M1.0  \\
s74928  &  53.176744747  &  -27.799670402  &  $18.51 \pm 0.000$  &  $2.04 \pm 0.004$  &  $1.55 \pm 0.001$  &  $0.59  \pm 0.000$  &  M4.0  \\
s79699  &  53.200369958  &  -27.78984038  &  $20.66 \pm 0.001$  &  $2.09 \pm 0.019$  &  $1.72 \pm 0.003$  &  $0.63  \pm 0.001$  &  M4.0  \\
s80618  &  53.162526096  &  -27.789645158  &  $24.48 \pm 0.009$  &  $1.83 \pm 0.120$  &  $1.24 \pm 0.023$  &  $0.40  \pm 0.014$  &  M2.0  \\
s82885  &  53.198434221  &  -27.784864101  &  $23.83 \pm 0.006$  &  $1.94 \pm 0.057$  &  $0.93 \pm 0.011$  &  $0.30  \pm 0.008$  &  M0.2  \\
s91263  &  53.148539616  &  -27.770133607  &  $22.49 \pm 0.002$  &  $1.86 \pm 0.016$  &  $0.94 \pm 0.004$  &  $0.27  \pm 0.003$  &  M0.9  \\
s92395  &  53.178155695  &  -27.769118708  &  $22.54 \pm 0.003$  &  $1.90 \pm 0.060$  &  $1.29 \pm 0.007$  &  $0.39  \pm 0.004$  &  M2.1  \\
s93532  &  53.197724177  &  -27.767764643  &  $24.07 \pm 0.006$  &  $2.12 \pm 0.142$  &  $1.48 \pm 0.020$  &  $0.45  \pm 0.010$  &  M2.0  \\
s93841  &  53.163849338  &  -27.76712288  &  $23.34 \pm 0.004$  &  $1.76 \pm 0.095$  &  $1.80 \pm 0.018$  &  $0.63  \pm 0.007$  &  M4.9  \\
s94206  &  53.13831609  &  -27.766346787  &  $22.37 \pm 0.002$  &  $2.34 \pm 0.095$  &  $1.84 \pm 0.010$  &  $0.63  \pm 0.004$  &  M4.3  \\
s95752  &  53.177677765  &  -27.763990756  &  $21.53 \pm 0.001$  &  $1.89 \pm 0.029$  &  $1.71 \pm 0.005$  &  $0.59  \pm 0.002$  &  M3.8  \\
s104030  &  53.131488485  &  -27.748464197  &  $21.16 \pm 0.001$  &  $1.79 \pm 0.007$  &  $0.81 \pm 0.001$  &  $0.18  \pm 0.001$  &  M0.0  \\
s104673  &  53.131060583  &  -27.743052772  &  $18.04 \pm 0.000$  &  $1.72 \pm 0.001$  &  $0.86 \pm 0.000$  &  $0.44  \pm 0.000$  &  M0.0  \\
s106781  &  53.153965475  &  -27.743636805  &  $23.53 \pm 0.004$  &  $1.86 \pm 0.152$  &  $2.01 \pm 0.026$  &  $0.70  \pm 0.008$  &  M6.0  \\
s107926  &  53.153218978  &  -27.741541723  &  $24.90 \pm 0.012$  &  $3.23 \pm 0.375$  &  $1.00 \pm 0.022$  &  $0.27  \pm 0.016$  &  M1.0  \\
s108066  &  53.146092506  &  -27.740987696  &  $24.38 \pm 0.008$  &  $2.94 \pm 0.366$  &  $1.50 \pm 0.025$  &  $0.48  \pm 0.012$  &  M3.0  \\
s108014  &  53.058106397  &  -27.734954191  &  $17.80 \pm 0.000$  &  $1.96 \pm 0.002$  &  $1.01 \pm 0.000$  &  $0.57  \pm 0.000$  &  M2.0  \\
s108497  &  53.093031113  &  -27.735720483  &  $19.42 \pm 0.000$  &  $1.97 \pm 0.004$  &  $1.16 \pm 0.001$  &  $0.36  \pm 0.000$  &  M1.7  \\
s110839  &  53.08965184  &  -27.732985693  &  $21.06 \pm 0.001$  &  $1.96 \pm 0.012$  &  $1.30 \pm 0.002$  &  $0.42  \pm 0.001$  &  M2.0  \\
s111269  &  53.113967436  &  -27.732641412  &  $22.58 \pm 0.002$  &  $2.10 \pm 0.049$  &  $1.40 \pm 0.007$  &  $0.47  \pm 0.004$  &  M3.0  \\
s111982  &  53.107924523  &  -27.728123879  &  $19.11 \pm 0.000$  &  $2.52 \pm 0.017$  &  $2.08 \pm 0.002$  &  $0.83  \pm 0.001$  &  M6.0  \\
s114688  &  53.086504259  &  -27.724249992  &  $22.22 \pm 0.002$  &  $1.84 \pm 0.018$  &  $0.89 \pm 0.004$  &  $0.27  \pm 0.003$  &  M0.0  \\
s115223  &  53.105133385  &  -27.724024515  &  $22.90 \pm 0.003$  &  $2.21 \pm 0.110$  &  $1.74 \pm 0.014$  &  $0.61  \pm 0.005$  &  M4.0  \\
s114563  &  53.069615748  &  -27.723369667  &  $19.90 \pm 0.000$  &  $2.37 \pm 0.017$  &  $1.98 \pm 0.002$  &  $0.79  \pm 0.001$  &  M6.0  \\
s116612  &  53.134679062  &  -27.721219304  &  $22.24 \pm 0.002$  &  $2.31 \pm 0.051$  &  $1.61 \pm 0.006$  &  $0.46  \pm 0.003$  &  M4.0  \\
s117391  &  53.087672048  &  -27.719481506  &  $21.04 \pm 0.001$  &  $1.97 \pm 0.011$  &  $1.09 \pm 0.002$  &  $0.38  \pm 0.001$  &  M1.3  \\
s119050  &  53.087850954  &  -27.717207134  &  $23.42 \pm 0.004$  &  $2.61 \pm 0.150$  &  $1.35 \pm 0.012$  &  $0.42  \pm 0.007$  &  M2.7  \\
s120814  &  53.11196534  &  -27.713255141  &  $22.19 \pm 0.002$  &  $2.00 \pm 0.060$  &  $1.80 \pm 0.009$  &  $0.68  \pm 0.003$  &  M4.2  \\
s123686  &  53.090461841  &  -27.706530349  &  $21.52 \pm 0.001$  &  $2.08 \pm 0.046$  &  $1.92 \pm 0.006$  &  $0.73  \pm 0.002$  &  M5.3  \\
s124539  &  53.077041221  &  -27.705926553  &  $23.22 \pm 0.004$  &  $2.06 \pm 0.347$  &  $2.53 \pm 0.052$  &  $1.05  \pm 0.009$  &  M8.7  \\
s124624  &  53.104858794  &  -27.705217424  &  $22.42 \pm 0.002$  &  $1.36 \pm 0.015$  &  $0.99 \pm 0.005$  &  $0.34  \pm 0.003$  &  M1.0  \\
s125478  &  53.100697649  &  -27.703049505  &  $21.29 \pm 0.001$  &  $2.74 \pm 0.343$  &  $3.08 \pm 0.026$  &  $1.24  \pm 0.003$  &  M9.0  \\
s126754  &  53.04936831  &  -27.701231055  &  $22.88 \pm 0.003$  &  $2.03 \pm 0.037$  &  $1.06 \pm 0.006$  &  $0.31  \pm 0.004$  &  M1.0  \\
s128173  &  53.06988597  &  -27.697118936  &  $21.57 \pm 0.001$  &  $2.12 \pm 0.029$  &  $1.55 \pm 0.004$  &  $0.54  \pm 0.002$  &  M3.3  \\
s128247  &  53.096591281  &  -27.694153422  &  $19.24 \pm 0.000$  &  $1.92 \pm 0.003$  &  $1.19 \pm 0.001$  &  $0.38  \pm 0.000$  &  M2.0  \\
s130804  &  53.060795882  &  -27.69126545  &  $22.35 \pm 0.002$  &  $1.88 \pm 0.020$  &  $0.98 \pm 0.004$  &  $0.31  \pm 0.003$  &  M1.0  \\
s132690  &  53.066065581  &  -27.687786007  &  $22.44 \pm 0.003$  &  $1.77 \pm 0.033$  &  $1.42 \pm 0.007$  &  $0.47  \pm 0.004$  &  M3.0  \\
s132844  &  53.048836128  &  -27.687739659  &  $23.68 \pm 0.006$  &  $1.66 \pm 0.046$  &  $0.63 \pm 0.008$  &  $0.12  \pm 0.008$  &  M4.4  \\
\enddata
\end{deluxetable}

\end{document}